\pdfoutput=1
\documentclass{llncs}
\usepackage{amsmath,amssymb,latexsym,amsfonts,amscd,stmaryrd,textcomp} 
\usepackage{mathtools}
\usepackage{mathrsfs}
\usepackage{bussproofs}

\usepackage[ruled,vlined,linesnumbered]{algorithm2e}

\usepackage{adjustbox}
\usepackage{booktabs}
\usepackage{subcaption}
\usepackage[color=blue!20]{todonotes}
\usepackage{verbatim}
\usepackage{enumerate}

\usepackage{apxproof}

\usepackage{wrapfig}

\usepackage{hyperref}

\usepackage{tikz}
\usetikzlibrary{backgrounds,arrows,shapes,positioning,fit,calc,decorations.pathmorphing,decorations.pathreplacing,trees}

\usepackage{comment}

\DefineNamedColor{named}{azul2}{RGB}{40,96,127}
\DefineNamedColor{named}{azul1}{RGB}{4,151,229}
\DefineNamedColor{named}{azul3}{RGB}{1,41,64}
\DefineNamedColor{named}{laranja1}{RGB}{255,146,5}

\newcommand{\sequence}{\mathbf{c}}

\renewcommand{\phi}{\varphi}
\newcommand{\error}{\mathit{error}}
\newcommand{\set}[1]{\ensuremath{\left\{#1\right\}}}
\newcommand{\tuple}[1]{\ensuremath{\left\langle#1\right\rangle}}

\newcommand{\size}[1]{\ensuremath{\left|#1\right|}} 

\newcommand{\csize}[1]{\ensuremath{\lceil #1\rceil}} 

\newcommand{\N}{\mathbb{N}}

\newcommand{\bigO}{\mathcal{O}}

\newcommand{\complexity}[1]{\textsc{#1}}

\newcommand{\PSpace}{\complexity{PSpace}}

\DeclareMathOperator{\dom}{dom}
\DeclareMathOperator{\img}{img}

\renewcommand{\vec}[1]{\mathbf{#1}}

\newcommand{\pfun}{\rightharpoonup}

\newcommand{\qedef}{\hfill$\triangle$}

\makeatletter
\newcommand{\oset}[3][0ex]{%
  \mathrel{\mathop{#3}\limits^{
    \vbox to#1{\kern-2\ex@
    \hbox{$\scriptstyle#2$}\vss}}}}
\makeatother

\newcommand{\SL}{\mathbf{SL}}

\newcommand{\true}{\mathsf{t}}

\renewcommand{\S}{s}
\renewcommand{\H}{h}
\newcommand{\SH}{(s,h)}
\newcommand{\SHpair}[2]{(#1,#2)}
\newcommand{\SHprime}{(s,h')}
\newcommand{\SHpprime}{(s',h')}
\newcommand{\SiHi}[1]{(s_{#1},h_{#1})}
\newcommand{\SHi}[1]{(s,h_{#1})}
\renewcommand{\SS}{\mathbf{S}}
\newcommand{\HH}{\mathbf{H}}
\newcommand{\Hoo}{\H_{1,1}} \newcommand{\Hot}{\H_{1,2}}
\newcommand{\Hto}{\H_{2,1}} \newcommand{\Htt}{\H_{2,2}}

\newcommand{\AbstListsKW}{\mathbf{AbstLists}}
\newcommand{\AbstLists}[2]{\AbstListsKW(#1,#2)}
\newcommand{\loc}{\ell}
\newcommand{\locs}{\mathsf{locs}}

\newcommand{\pto}{\mapsto}

\newcommand{\emp}{\mathbf{emp}}

\newcommand{\atomic}{\tau}

\newcommand{\nil}{\mathsf{nil}}

\newcommand{\Var}{\mathbf{Var}}
\newcommand{\Loc}{\mathbf{Loc}}

\renewcommand{\qedhere}{\qed}

\newcommand{\mw}{{-\!\!\ast}}
\newcommand{\sept}{{-\!\!\circledast}}
\newcommand{\IteratedStar}{\mathop{\scalebox{1.5}{\raisebox{-0.2ex}{$\ast$}}}}%

\newcommand{\RuleName}[1]{\mathtt{#1}}
\newcommand{\ls}{\RuleName{ls}}

\newcommand{\stdunion}{+}

\newcommand{\oursunion}
{\uplus^{\S}}
\newcommand{\oursunionop}[1]
{\uplus^{#1}}
\newcommand{\oursup}{\scriptscriptstyle\mathsf{st}}
\newcommand{\stdsup}{~\scriptscriptstyle\mathsf{wk}}
\newcommand{\imodels}[1]{\oset[-0.6ex]{#1}{\models}}
\newcommand{\notimodels}[1]{\oset[-0.3ex]{#1}{\not\models}}
\newcommand{\stdmodels}{\imodels{\stdsup}}
\newcommand{\ourmodels}{\imodels{\oursup}}
\newcommand{\nourmodels}{\notimodels{\oursup}}
\newcommand{\anymodels}{\models}
\newcommand{\ientails}[1]{\imodels{\oursup}_{#1}}

\newcommand{\entailsx}{\ientails{\vec{x}}}

\newcommand{\fvs}[1]{\mathsf{fvs}(#1)}

\newcommand{\EX}[1]{\exists #1\ldotp}
\newcommand{\FA}[1]{\forall #1\ldotp}

\newcommand{\Iso}{\cong}

\newcommand{\alloc}[1]{\mathbf{alloc}(#1)}

\newcommand{\allocp}[1]{\mathsf{alloc}(#1)}

\newcommand{\assert}[1]{\set{#1}}
\newcommand{\triple}[3]{\assert{#1}#2\assert{#3}}
\newcommand{\istmt}[1]{\quad\mathtt{#1}}

\newcommand{\modvars}[1]{\mathsf{modifiedVars}(#1)}

\newcommand{\tripl}[3]{\set{#1} #2 \set{#3}}
\newcommand{\fnext}{\mathsf{next}}
\newcommand{\progcmt}[1]{\quad\text{\texttt{\# #1}}}

\newcommand{\eqone}{=\!1}
\newcommand{\geqtwo}{\geq\!2}
\newcommand{\theListInts}{\set{\eqone,\geqtwo}}

\newcommand{\A}{\mathcal{A}}
\newcommand{\B}{\mathcal{B}}

\newcommand{\V}{V}
\newcommand{\E}{E}

\newcommand{\garb}{\gamma}

\newcommand{\Compose}{\bullet}
\newcommand{\isofun}{\sigma}

\newcommand{\gmw}{{-\!\!\bullet}}
\newcommand{\gsept}{{-\!\!\bullet}}

\newcommand{\iedges}{\mathsf{edges}}
\newcommand{\abst}[2]{\mathsf{abst}_{#1}(#2)}

\newcommand{\subheaps}{\mathsf{subHeaps}}

\newcommand{\chunks}{\mathsf{chunks}}
\newcommand{\goodchunksKW}{\chunks^{+}}
\newcommand{\goodchunks}{\goodchunksKW}
\newcommand{\badchunksKW}{\chunks^{-}}
\newcommand{\badchunks}{\badchunksKW}

\newcommand{\mfalloc}{\mathsf{alloc}^{-}}

\newcommand{\Rset}{\rho}

\newcommand{\mabst}{\mathsf{ams}}

\newcommand{\mabsti}[1]{\mathsf{abst}_{#1}}

\newcommand{\AMS}{\mathbf{AMS}}

\newcommand{\AMSks}{\mathbf{AMS}_{k,\S}}

\newcommand{\Aset}{\mathbf{A}}
\newcommand{\lift}[3]{\mathsf{lift}_{#1 \nearrow #2}(#3)}

\newcommand{\normalform}[1]{\mathsf{NF}_{\vec{x}}(#1)}

\newcommand{\formofxm}[1]{\mathsf{AMS2SL}^{m}(#1)}

\newcommand{\formof}[2]{\mathsf{AMS2SL}^{#1}(#2)}

\newcommand{\qbftr}[1]{\mathsf{qbf\_to\_sl}(#1)}
\newcommand{\qbfaux}[1]{\mathsf{aux}(#1)}

\tikzset{loc/.style={draw,rounded corners,minimum size=14pt,inner
    sep=2pt}}
    
\tikzset{dangling/.style={fill=orange!50}}

\tikzset{hnode/.style={draw,circle,minimum size=14pt,inner
    sep=2pt}}
\tikzset{stable/.style={fill=blue!20}}
\tikzset{hedge/.style={->}}

\tikzset{rgnode/.style={draw,thick,rounded corners,minimum size=14pt,inner
    sep=2pt}}
\tikzset{rgedge/.style={->,thick}}

\newcommand{\convexpath}[2]{
[
    create hullnodes/.code={
        \global\edef\namelist{#1}
        \foreach [count=\counter] \nodename in \namelist {
            \global\edef\numberofnodes{\counter}
            \node at (\nodename) [draw=none,name=hullnode\counter] {};
        }
        \node at (hullnode\numberofnodes) [name=hullnode0,draw=none] {};
        \pgfmathtruncatemacro\lastnumber{\numberofnodes+1}
        \node at (hullnode1) [name=hullnode\lastnumber,draw=none] {};
    },
    create hullnodes
]
($(hullnode1)!#2!-90:(hullnode0)$)
\foreach [
    evaluate=\currentnode as \previousnode using \currentnode-1,
    evaluate=\currentnode as \nextnode using \currentnode+1
    ] \currentnode in {1,...,\numberofnodes} {
-- ($(hullnode\currentnode)!#2!-90:(hullnode\previousnode)$)
  let \p1 = ($(hullnode\currentnode)!#2!-90:(hullnode\previousnode) - (hullnode\currentnode)$),
    \n1 = {atan2(\y1,\x1)},
    \p2 = ($(hullnode\currentnode)!#2!90:(hullnode\nextnode) - (hullnode\currentnode)$),
    \n2 = {atan2(\y2,\x2)},
    \n{delta} = {-Mod(\n1-\n2,360)}
  in
    {arc [start angle=\n1, delta angle=\n{delta}, radius=#2]}
}
-- cycle
}

\newcommand{\tikzAMS}[3]{
  \begin{tikzpicture}
    \node[draw,rectangle,azul1] (rg) {#1};
    \node[draw,rectangle,laranja1,below=5mm of rg.south west,anchor=west,dashed] (Rset) {#2};
    \node[draw,laranja1,rectangle,right=1mm of Rset,dashed] (garb) {#3};
    \begin{scope}[on background layer]
      \node[fit=(rg)(Rset)(garb), fill=gray!10] {};
    \end{scope}

  \end{tikzpicture}
}

\includecomment{AppendixOnly}
\excludecomment{SubmissionOnly}

\title{Strong-Separation Logic}

\author{Jens Pagel \and Florian Zuleger}
\institute{TU Wien, Austria\\
\email{pagel@forsyte.at, zuleger@forsyte.at}
}

\begin{document}

\maketitle

\begin{abstract}
  Most automated verifiers for separation logic are based on the symbolic-heap fragment, which disallows both the magic-wand operator and the application of classical Boolean operators to spatial formulas.
  This is not surprising, as support for the magic wand quickly leads to undecidability, especially when combined with inductive predicates for reasoning about data structures.
  To circumvent these undecidability results, we propose assigning a more restrictive semantics to the separating conjunction.
  We argue that the resulting logic, strong-separation logic, can be used for symbolic execution and abductive reasoning just like ``standard'' separation logic, while remaining decidable even in the presence of both the magic wand and the list-segment predicate---a combination of features that leads to undecidability for the standard semantics.
\end{abstract}

\section{Introduction}\label{sec:introduction}
Separation logic~\cite{reynolds2002separation} is one of the most
successful formalisms for the analysis and verification of programs
making use of dynamic resources such as heap memory and access
permissions~\cite{bornat2005permission,ohearn2007resources,calcagno2011compositional,berdine2011slayer,dudka2011predator,jacobs2011verifast,calcagno2015moving}.
At the heart of the success of separation logic (SL) is the \emph{separating conjunction}, $*$, which supports concise statements about the disjointness of resources.
In this article, we will focus on separation logic for describing the heap in single-threaded heap-manipulating programs.
In this setting, the formula $\phi*\psi$ can be read as ``the heap can be split into two disjoint parts, such that $\phi$ holds for one part and $\psi$ for the other.''

Our article starts from the following observation: The standard
semantics of $*$ allows splitting a heap into two arbitrary sub-heaps.
The magic-wand operator $\mw$, which is the adjoint of $*$, then
allows adding arbitrary heaps.  This arbitrary splitting and adding of
heaps makes reasoning about SL formulas difficult, and quickly renders
separation logic undecidable when inductive predicates for data
structures are considered.  For example, Demri et al.~recently showed
that adding only the singly-linked list-segment predicate to
propositional separation logic (i.e., with $*,\mw$ and classical
connectives $\wedge,\vee,\neg$) leads to
undecidability~\cite{demri2018effects}.

Most SL specifications used
in automated verification do not, however, make use of arbitrary
heap compositions.
For example, the widely used symbolic-heap fragments of separation
logic considered, e.g.,
in~\cite{berdine2004decidable,berdine2005symbolic,cook2011tractable,iosif2013tree,iosif2014deciding},
have the following property:
a symbolic heap satisfies a separating conjunction, if and only if one can split the model at locations that are the values of some program variables.

Motivated by this observation, we propose a more restrictive
separating conjunction that allows splitting the heap only at location that are the values of some program variables.
We call the resulting logic \emph{strong-separation logic}.
Strong-separation logic (SSL) shares many properties with standard
separation-logic semantics; for example, the models of our logic form
a separation algebra.
Because the \emph{frame rule} and other standard SL inference rules
continue to hold for SSL, SSL is suitable for deductive Hoare-style
verification \`{a} la~\cite{ishtiaq01bi,reynolds2002separation},
symbolic execution~\cite{berdine2005symbolic}, as well as 
abductive reasoning ~\cite{calcagno2011compositional,calcagno2015moving}.
At the same time, SSL has much better computational properties than standard SL---especially when formulas contain expressive features such as the \emph{magic wand}, $\mw$, or negation.

We now give a more detailed introduction to the contributions of this article.

\vspace{-0.2cm}
\paragraph*{The standard semantics of the separating conjunction.}
To be able to justify our changed semantics of $*$, we need to introduce a bit of terminology.
As standard in separation logic, we interpret SL formulas over \emph{stack--heap pairs}.
A \emph{stack} is a mapping of the program variables to memory locations.
A \emph{heap} is a finite partial function between memory locations;
if a memory location $l$ is mapped to location $l'$, we say the heap contains a \emph{pointer} from $l$ to $l'$.  A memory location $l$ is \emph{allocated} if there is a pointer of the heap from $l$ to some location $l'$.
We call a location \emph{dangling} if it is the target of a pointer but not allocated;
a pointer is dangling if its target location is dangling.

Dangling pointers arise naturally in compositional specifications, i.e., in formulas that employ the separating conjunction $*$:
In the standard semantics of separation logic, a stack--heap pair $\SH$ satisfies a formula $\phi*\psi$, if it is possible to split the heap $\H$ into two disjoint parts $\H_1$ and $\H_2$ such that $\SHpair{\S}{\H_1}$ satisfies $\phi$ and $\SHpair{\S}{\H_2}$ satisfies
$\psi$.
Here, disjoint means that the allocated locations of $\H_1$ and $\H_2$ are disjoint;
however, the targets of the pointers of $\H_1$ and $\H_2$ do not have to be disjoint.

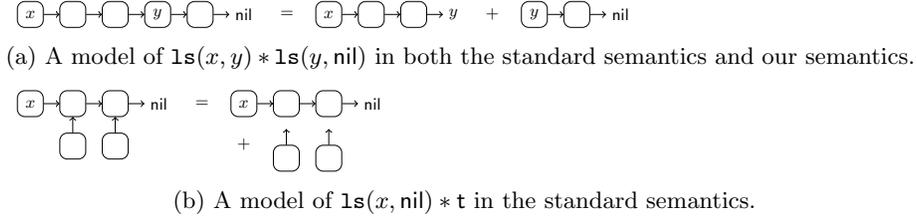
\begin{figure}[tb!]
  \newcommand{\ndist}{3mm}
\newcommand{\ListNonDangling}{
\begin{tikzpicture}
  \node[loc] (x) {$x$};
  \node[loc,right=\ndist of x] (x1) {};
  \node[loc,right=\ndist of x1] (x2) {};
\node[loc,right=\ndist of x2] (y) {$y$};
\node[loc,right=\ndist of y] (y1) {};
\node[right=\ndist of y1] (nil) {$\nil$};
\draw (x) edge[->] (x1);
\draw (x1) edge[->] (x2);
\draw (x2) edge[->] (y);
\draw (y) edge[->] (y1);
\draw (y1) edge[->] (nil);

\node[right=\ndist of nil] (eq) {$=$};

  \node[loc,right=\ndist of eq] (ax) {$x$};
  \node[loc,right=\ndist of ax] (ax1) {};
  \node[loc,right=\ndist of ax1] (ax2) {};
\node[right=\ndist of ax2] (ay) {$y$};

\node[right=\ndist of ay] (stdunion) {$\stdunion$};

\node[loc,right=\ndist of stdunion] (ayprime) {$y$};
\node[loc,right=\ndist of ayprime] (ay1) {};
\node[right=\ndist of ay1] (anil) {$\nil$};
\draw (ax) edge[->] (ax1);
\draw (ax1) edge[->] (ax2);
\draw (ax2) edge[->] (ay);
\draw (ayprime) edge[->] (ay1);
\draw (ay1) edge[->] (anil);
\end{tikzpicture}
}

\newcommand{\ListStarTrue}{
\begin{tikzpicture}
  \node[loc] (x) {$x$};
  \node[loc,right=\ndist of x] (x1) {};
  \node[loc,right=\ndist of x1] (x2) {};
  \node[right=\ndist of x2] (nil) {$\nil$};

  \node[loc,below=\ndist of x1] (d1) {};
  \node[loc,below=\ndist of x2] (d2) {};

  \node[right=\ndist of nil] (eq) {$=$};

\draw (x) edge[->] (x1);
\draw (x1) edge[->] (x2);
\draw (x2) edge[->] (nil);
\draw (d1) edge[->] (x1);
\draw (d2) edge[->] (x2);

  \node[loc,right=\ndist of eq] (ax) {$x$};
  \node[loc,right=\ndist of ax] (ax1) {};
  \node[loc,right=\ndist of ax1] (ax2) {};
  \node[right=\ndist of ax2] (anil) {$\nil$};

  \node[below=\ndist of ax] (stdunion) {$\stdunion$};
  \node[below=\ndist of ax1,yshift=0.3cm] (bx1) {};
  \node[below=\ndist of ax2,yshift=0.3cm] (bx2) {};
  \node[loc,below=\ndist of bx1] (bd1) {};
  \node[loc,below=\ndist of bx2] (bd2) {};

\draw (ax) edge[->] (ax1);
\draw (ax1) edge[->] (ax2);
\draw (ax2) edge[->] (anil);
\draw (bd1) edge[->] (bx1);
\draw (bd2) edge[->] (bx2);
\end{tikzpicture}
}

  \centering
  \begin{subfigure}{\columnwidth}
    \scalebox{0.7}{
      \ListNonDangling
      }
    \caption{A model of $\ls(x,y)*\ls(y,\nil)$ in both the standard semantics and our semantics.
    \vspace{0.2cm}}
    \label{fig:dangling-pointers:ls}
  \end{subfigure}
  \begin{subfigure}{\columnwidth}
    \scalebox{0.7}{
      \ListStarTrue
      }
    \caption{A model of $\ls(x,\nil)*\true$ in the standard semantics.}\label{fig:dangling-pointers:true}
  \end{subfigure}
  \caption{Two models and their decomposition into disjoint submodels.
  Dangling arrows represent dangling pointers.
    \vspace{-0.5cm}}
  \label{fig:dangling-pointers}
\end{figure}

We illustrate
this in Fig.~\ref{fig:dangling-pointers:ls}, where we show a graphical representation of a stack--heap pair $\SH$ that satisfies the formula $\ls(x,y)*\ls(y,\nil)$.
Here, $\ls$ denotes the list-segment predicate.
As shown in Fig.~\ref{fig:dangling-pointers:ls}, $\H$ can be split into two disjoint parts $\H_1$ and $\H_2$ such that $\SHpair{\S}{\H_1}$ is a model of $\ls(x,y)$ and $\SHpair{\S}{\H_2}$ is a model of $\ls(y,\nil)$.
Now, $\H_1$ has a dangling pointer with target $\S(y)$ (displayed with an orange background), while no pointer is dangling in the heap $\H$.

\vspace{-0.2cm}
\paragraph*{In what sense is the standard semantics too permissive?}
The standard semantics of $*$ allows splitting a heap into two
arbitrary sub-heaps, which may result in the introduction of arbitrary
dangling pointers into the sub-heaps.
We note, however, that the introduction of dangling pointers is
\emph{not} arbitrary when splitting the models of
$\ls(x,y)*\ls(y,\nil)$; there is only one way of splitting the models
of this formula, namely at the location of program variable $y$.
The formula $\ls(x,y)*\ls(y,\nil)$ belongs to a certain variant of the
symbolic-heap fragment of separation logic, and all
formulas of this fragment have the property that their models can only
be split at locations that are the values of some variables.

Standard SL semantics also allows the introduction of dangling
pointers without the use of variables.
Fig.~\ref{fig:dangling-pointers:true} shows a model of
$\ls(x,\nil) * \true$---assuming the standard semantics. Here, the
formula $\true$ (for \emph{true}) stands for any arbitrary heap. In
particular, this includes heaps with arbitrary dangling pointers into
the list segment $\ls(x,\nil)$.  This power of introducing arbitrary
dangling pointers is what is used by Demri et al.~for their
undecidability proof of propositional separation logic with the
singly-linked list-segment predicate~\cite{demri2018effects}.

\vspace{-0.2cm}
\paragraph*{Strong-separation logic.}
In this article, we want to explicitly \emph{disallow} the \emph{implicit} sharing of dangling locations when composing heaps.
We propose to parameterize the separating conjunction by the stack and
exclusively allow the union of heaps that only share locations that are pointed to by the stack.
For example, the model in Fig.~\ref{fig:dangling-pointers:true} is
\emph{not} a model of $\ls(x,\nil)*\true$ in our semantics because of the dangling pointers in the sub-heap that satisfies $\true$.
\emph{Strong-separation logic} (SSL) is the logic resulting from this restricted definition of the separating conjunction.

\vspace{-0.2cm}
\paragraph*{Why should I care?}
We argue that SSL is a promising proposal for automated program verification:

1) We show that the memory models of strong-separation logic form a \emph{separation algebra}~\cite{calcagno2007local}, which guarantees
the soundness of the standard \emph{frame rule} of SL~\cite{reynolds2002separation}.
Consequently, SSL can be potentially be used instead of standard SL in a wide variety of (semi-)automated analyzers and verifiers, including Hoare-style verification~\cite{reynolds2002separation}, symbolic execution~\cite{berdine2005symbolic}, and bi-abductive shape
analysis~\cite{calcagno2011compositional}.


2) To date, most automated reasoners for separation logic have been developed for \emph{symbolic-heap separation
  logic}~\cite{berdine2004decidable,berdine2005symbolic,calcagno2011compositional,iosif2013tree,iosif2014deciding,katelaan2019effective,journals/corr/abs-2002-01202,conf/lpar/KatelaanZ20}.
In these fragments of separation logic, assertions about the heap can exclusively be combined via $*$; neither the magic wand $\mw$ nor classical Boolean connectives are permitted.
We show
that the strong semantics agrees with the standard semantics on symbolic heaps.
For this reason, symbolic-heap SL specifications remain unchanged when switching to strong-separation logic.

3) We establish that the satisfiability and entailment problem for full propositional separation logic with the singly-linked list-segment predicate is decidable in our semantics (in $\PSpace$)---in stark contrast to the aforementioned undecidability result obtained by Demri et al.~\cite{demri2018effects} assuming the standard
semantics.

4) The standard Hoare-style approach to verification requires discharging verification conditions (VCs), which amounts to proving for loop-free pieces of code that a pre-condition implies some post-condition.
Discharging VCs can be automated by calculi that symbolically execute the pre-condition forward resp. the post-condition backward, and then using an entailment checker for proving the implication.
For SL, symbolic execution calculi can be formulated using the magic wand resp. the septraction operator.
However, these operators have proven to be difficult for automated procedures: ``VC-generators do not work especially well with separation logic, as they introduce magic-wand $\mw$ operators which are difficult to eliminate.''~\cite[p.~131]{appel2014program}
In contrast, we demonstrate that SSL can overcome the described difficulties.
We formulate a forward symbolic execution calculus for a simple heap-manipulating programming language using SSL.
In conjunction with our entailment checker, see 3), our calculus gives rise to a fully-automated procedure for discharging VCs of loop-free code segments.

5) Computing solutions to the \emph{abduction problem} is an integral building block of Facebook's Infer
analyzer~\cite{calcagno2015moving}, required for a scalable and fully-automated shape analysis~\cite{calcagno2011compositional}.
We show how to compute explicit representations of optimal, i.e., \emph{logically weakest} and \emph{spatially minimal}, solutions to the abduction problem for the separation logic considered in this paper.
The result is of theoretical interest, as explicit representations for optimal solutions to the abduction problem are hard to obtain~\cite{calcagno2011compositional,gorogiannis2011complexity}.

\paragraph*{Contributions.}
Our main contributions are as follows:
\begin{enumerate}
\item We propose and motivate \emph{strong-separation logic} (SSL), a new semantics for
  separation logic.
\item We present a $\PSpace$ decision procedure for strong-separation logic with points-to assertions, the list-segment predicate $\ls(x,y)$, and spatial and classical operators, i.e., $*,\mw,\wedge,\vee,\neg$---a logic that is undecidable when assuming the standard semantics~\cite{demri2018effects}.
\item We present symbolic execution rules for SSL, which allow us to discharge verification conditions fully automatically.
\item We show how to compute explicit representations of optimal solutions to the abduction problem for the SSL considered in (2).
  \end{enumerate}
  We strongly believe that these results motivate further research on SSL (e.g., going beyond the singly-linked list-segment predicate, implementing our decision procedure and integrating it into fully-automated analyzers).

\paragraph*{Related work.}

The undecidability of separation logic was established already in~\cite{calcagno2001computability}.
Since then, decision problems for a large number of fragments and variants of separation logic have been studied.
Most of this work has been on symbolic-heap separation logic or other variants of the logic that neither support the magic wand nor the use of negation below the $*$ operator.
While entailment in the symbolic-heap fragment with inductive definitions is undecidable in general~\cite{antonopoulos2014foundations}, there are decision procedures for variants with built-in lists and/or trees~\cite{berdine2004decidable,cook2011tractable,navarro2013separation,piskac2013automating,piskac2014automating},
support for defining variants of linear structures~\cite{gu2016complete} or tree structures~\cite{tatsuta2015separation,iosif2014deciding} or graphs of bounded tree width~\cite{iosif2013tree,katelaan2019effective}.
The expressive heap logics \textsc{Strand}~\cite{madhusudan2011decidable} and
\textsc{Dryad}~\cite{qiu2013natural} also have decidable fragments, as have some other separation logics that allow combining shape and data constraints.
Besides the already mentioned work~\cite{piskac2013automating,piskac2014automating}, these include~\cite{le2017decidable,katelaan2018separation}.

Among the aforementioned works, the graph-based decision procedures of~\cite{cook2011tractable} and~\cite{katelaan2018separation} are most closely related to our approach.
Note however, that neither of these works supports reasoning about magic wands or negation below the separating conjunction.

In contrast to symbolic-heap SL, separation logics with the \emph{magic wand} quickly become undecidable.  Propositional separation logic with the magic wand, but without inductive data structures, was shown to be decidable in $\PSpace$ in the early days of SL research~\cite{calcagno2001computability}.
Support for this fragment was added to CVC4 a few years
ago~\cite{reynolds2016decision}.
Some tools have ``lightweight'' support for the magic wand involving heuristics and user annotations, in part motivated by the lack of
decision procedures~\cite{blom2015witnessing,schwerhoff2015lightweight}.

There is a significant body of work studying first-order SL with the magic wand and unary points-to assertions, but without a list predicate.
This logic was first shown to be undecidable in~\cite{brochenin2012almighty};
a result that has since been refined, showing e.g.~that while satisfiability is still in $\PSpace$ if we allow one quantified variable~\cite{demri2014separation}, two variables already lead to undecidability, even without the separating conjunction~\cite{demri2014expressive}.
Echenim et al.~\cite{echenim2019bernays} have recently addressed the
satisfiability problem of SL with $\exists^{*}\forall^{*}$ quantifier prefix, separating conjunction, magic wand, and full Boolean closure, but no inductive definitions.
The logic was shown to be undecidable in general (contradicting an earlier claim~\cite{reynolds2017reasoning}),
but decidable in $\PSpace$ under certain restrictions.

\paragraph*{Outline.}
In Section~\ref{sec:separation}, we introduce two semantics of propositional separation logic, the standard semantics and our new \emph{strong-separation} semantics. 
We show the decidability of the satisfiability and entailment problems of SSL with lists in Section~\ref{sec:deciding}.
We present symbolic execution rules for SSL in Section~\ref{sec:verification}.
We show how to compute explicit representations of optimal solutions to the abduction problem in Section~\ref{sec:abduction}.
We conclude in Section~\ref{sec:conclusion}.
All missing proofs are given in the extended version~\cite{journals/corr/abs-2001-06235} for space reasons.


\section{Strong- and Weak-Separation Logic}\label{sec:separation}

\subsection{Preliminaries}

We denote by $\size{X}$ the cardinality of the set $X$.
Let $f$ be a (partial) function.
Then, $\dom(f)$ and $\img(f)$ denote the domain and image of $f$, respectively.
We write $\size{f} := \size{\dom(f)}$ and $f(x) = \bot$ for $x \not\in \dom(f)$.
We frequently use set notation to define and reason about partial functions:
$f := \set{x_1 \mapsto y_1, \ldots, x_k \mapsto y_k}$ is the partial function that maps $x_i$ to $y_i$, $1 \leq i \leq k$, and is undefined on all other values;
$f^{-1}(b)$ is the set of all elements $a$ with $f(a) = b$;
we write $f \cup g$ resp. $f \cap g$ for the union resp. intersection of partial functions $f$ and $g$, provided that $f(a)=g(a)$ for all $a \in \dom(f) \cap \dom(g)$;
similarly, $f \subseteq g$ holds if $\dom(f)\subseteq\dom(g)$.
Given a partial function $f$, we denote by $f[x/v]$ the updated partial function in which $x$ maps to $v$, i.e.,
$$f[x/v](y) = \begin{cases}
               v, & \text{if } y=x, \\
               f(y) & \text{otherwise},
             \end{cases}$$
where we use $v = \bot$ to express that the updated function $f[x/v]$ is undefined for $x$.

Sets and ordered sequences are denoted in boldface, e.g.,
$\vec{x}$.
To list the elements of a sequence, we write $\tuple{x_1,\ldots,x_k}$.

We assume a linearly-ordered infinite set of variables $\Var$ with $\nil \in \Var$ and denote by $\max(\vec{v})$ the maximal variable among a set of variables $\vec{v}$ according to this order.
In Fig.~\ref{fig:syntax}, we define the syntax of the separation-logic fragment we study in this article.
The atomic formulas of our logic are the \emph{empty-heap predicate} $\emp$, \emph{points-to assertions} $x \pto y$, the \emph{list-segment predicate} $\ls(x,y)$, equalities $x = y$ and disequalities $x \neq y$\footnote{As our logic contains negation, $x \neq y$ can be expressed as $\neg (x = y)$. However, we treat disequalities as atomic to be able to use them in the positive fragment of our logic, defined later, which precludes the use of negation. }
; in all these cases, $x,y \in\Var$.
Formulas are closed under the classical Boolean operators
$\wedge,\vee,\neg$ as well as under the \emph{separating conjunction} $*$ and the existential magic wand, also called \emph{septraction}, $\sept$ (see e.g.~\cite{brochenin2012almighty}).
We collect the set of all SL formulas in $\SL$.
We also consider derived operators and formulas, in particular the \emph{separating implication} (or \emph{magic wand}), $\mw$, defined by $\phi \mw \psi := \neg(\phi \sept \neg \psi)$.\footnote{As $\mw$ can be defined via $\sept$ and $\neg$ and vice-versa, the expressivity of our logic does not depend on which operator we choose.
We have chosen $\sept$ because we can include this operator in the positive fragment considered later on.}
We also use \emph{true}, defined as $\true := \emp \vee \neg \emp$.
Finally, for $\Phi=\set{\phi_1,\ldots,\phi_n}$, we set $\IteratedStar\Phi := \phi_1 * \phi_2 * \cdots * \phi_n$, if $n > 1$,
and $\IteratedStar\Phi := \emp$, if $n = 0$.
By $\fvs{\phi}$ we denote the set of (free) variables of $\phi$.
We define the \emph{size} of the formula $\phi$ as $\size{\phi} = 1$
for atomic formulas $\phi$,
$\size{\phi_1 \times \phi_2} := \size{\phi_1}+\size{\phi_2}+1$ for
$\times \in \set{\wedge,\vee,*,\sept}$ and
$\size{\neg\phi_1}:=\size{\phi_1}+1$.

\begin{figure}[tb!]
  \centering
  \input{tablesandfigures/syntax}
  \vspace{-0.5cm}
  \caption{The syntax of separation logic with list segments.
  \vspace{-0.5cm}
  }
  \label{fig:syntax}
\end{figure}
\begin{figure*}[tb!]
   \centering
    \input{tablesandfigures/semantics}
  \vspace{-0.5cm}
  \caption{The standard, ``weak'' semantics of separation logic, $\stdmodels$, and the ``strong'' semantics, $\ourmodels$.
  We write $\anymodels$ when there is no difference between $\stdmodels$ and $\ourmodels$.
  \vspace{-0.5cm}
  }
  \label{fig:semantics}
\end{figure*}
\subsection{Two Semantics of Separation Logic}

\paragraph*{Memory model.}
$\Loc$ is an infinite set of \emph{heap locations}. 
A \emph{stack} is a partial function $\S\colon \Var \pfun \Loc$.
A \emph{heap} is a partial function $\H\colon \Loc \pfun \Loc$. 
A \emph{model} is a stack--heap pair $\SH$ with $\nil\in\dom(\S)$ and
$\S(\nil) \notin\dom(\H)$.
We let $\locs(\H) := \dom(\H) \cup \img(\H)$.
A location $\loc$ is \emph{dangling} if
$\loc \in \img(\H)\setminus\dom(\H)$.
We write $\SS$ for the set of all stacks and $\HH$ for the set of all heaps.

\paragraph{Two notions of disjoint union of heaps.}
We write $\H_1 \stdunion \H_2$ for the union of disjoint heaps, i.e.,
\vspace{-0.5cm}
\[
  \H_1 \stdunion \H_2 :=
  \begin{cases}
    \H_1 \cup \H_2, & \text{if } \dom(\H_1)\cap\dom(\H_2)=\emptyset\\
    \bot,& \text{otherwise.}
    \vspace{-0.2cm}
  \end{cases}
\]
This standard notion of disjoint union is commonly used to assign semantics to the separating conjunction and magic wand.
It requires that $\H_1$ and $\H_2$ are domain-disjoint, but does not impose any restrictions on the \emph{images} of the heaps.
In particular, the dangling pointers of $\H_1$ may alias arbitrarily with the domain and image of $\H_2$ and vice-versa.

Let $\S$ be a stack.
We write $\H_1 \oursunion \H_2$ for the disjoint union of $\H_1$ and $\H_2$ that restricts aliasing of dangling pointers to the locations in stack $\S$.
This yields an infinite family of union operators:
one for each stack.
Formally,
\vspace{-0.2cm}
\[\H_1 \oursunion \H_2 :=
  \begin{cases}
    \H_1 \stdunion \H_2,& \text{if } \locs(\H_1)\cap\locs(\H_2)
    \subseteq \img(\S) \\
    \bot, & \text{otherwise.}
    \vspace{-0.2cm}
  \end{cases}
\]
Intuitively, $\H_1\oursunion\H_2$ is the (disjoint) union of heaps that share only locations that are in the image of the stack $\S$.
Note that if $\H_1\oursunion\H_2$ is defined then  $\H_1\stdunion\H_2$ is defined, but not vice-versa.

Just like the standard disjoint union $\stdunion$, the operator $\oursunion$ gives rise to a separation algebra, i.e., a cancellative, commutative partial  monoid~\cite{calcagno2007local}:

\begin{lemma}\label{lem:separation-algebra}
  Let $\S$ be a stack and let $u$ be the empty heap (i.e., $\dom(u)=\emptyset$).
  The triple $(\HH,\oursunion,u)$ is a separation algebra.
\end{lemma}
\begin{AppendixOnly}
\begin{appendixproof}
  \emph{(Lemma~\ref{lem:separation-algebra}.)}
  Trivially, the operation $\oursunion$ is commutative and associative with unit $u$.
  Let $\H \in \HH$.
  Consider $\H_1, \H_2 \in \HH$ such that $\H \oursunion \H_1 = \H \oursunion \H_2 \neq \bot$.
  Since the domain of $\H$ is disjoint from the domains of $\H_1$ and $\H_2$, it follows that for all $x$, $\H_1(x)=\H_2(x)$ and thus $\H_1=\H_2$.
  As $\H_1$ and $\H_2$ were chosen arbitrarily, we obtain that the function $\H \oursunion (\cdot)$ is injective. Consequently, the monoid is cancellative. 
\end{appendixproof}
\end{AppendixOnly}

\paragraph*{Weak- and strong-separation logic.}
Both $\stdunion$ and $\oursunion$ can be used to give a semantics to the separating conjunction and septraction. We denote the corresponding model relations $\stdmodels$ and $\ourmodels$ and define them in Fig.~\ref{fig:semantics}.
Where the two semantics agree, we simply write $\anymodels$.

In both semantics, $\emp$ only holds for the empty heap, and $x=y$ holds for the empty heap when $x$ and $y$ are interpreted by the same location\footnote{Usually $x=y$ is defined to hold for \emph{all} heaps, not just the empty heap, when $x$ and $y$ are interpreted by the same location;
however, this choice does not change the expressivity of the logic: the formula $(x=y) * \true$ expresses the standard semantics.
Our choice is needed for the results on the positive fragment considered in Section~\ref{sec:correspondence}}.
Points-to assertions $x\pto y$ are precise, i.e., only hold in singleton heaps.
(It is, of course, possible to express intuitionistic points-to assertions by $x \pto y * \true$.)
The list segment predicate $\ls(x,y)$ holds in possibly-empty lists of pointers from $\S(x)$ to $\S(y)$. 
The semantics of Boolean connectives are standard.
The semantics of the separating conjunction, $*$, and septraction, $\sept$, differ based on the choice of $\stdunion$ vs.~$\oursunion$ for combining disjoint heaps.
In the former case, denoted $\stdmodels$, we get the standard semantics of separation logic (cf.~\cite{reynolds2002separation}).
In the latter case, denoted $\ourmodels$, we get a semantics that imposes stronger requirements on sub-heap composition:
Sub-heaps may only overlap at locations that are stored in the stack.

Because the semantics $\ourmodels$ imposes stronger constraints, we will refer to the standard semantics $\stdmodels$ as the \emph{weak} semantics of separation logic and to the semantics $\ourmodels$ as the \emph{strong} semantics of separation logic.
Moreover, we use the terms \emph{weak-separation logic} (WSL) and \emph{strong-separation logic} (SSL) to distinguish between SL with the semantics $\stdmodels$ and $\ourmodels$.

  \begin{figure}[tb!]
    \centering
    \newcommand{\ndist}{5mm}
\begin{tabular}{l|ll}
\begin{minipage}[b]{0.35\linewidth}
\begin{tikzpicture}
  \node[loc] (x) {$a$};
  \node[loc,right=\ndist of x] (x1) {};
  \node[loc,right=\ndist of x1] (x2) {};
\node[right=\ndist of x2] (y) {$\nil$};
\node[loc,below=\ndist of x] (b) {$b$};
\node[right=\ndist of b] (nil) {$\nil$};
\draw (x) edge[->] (x1);
\draw (x1) edge[->] (x2);
\draw (x2) edge[->] (y);
\draw (b) edge[->] (nil);
\end{tikzpicture}
\end{minipage}
&
$\quad\quad$
&
\begin{minipage}[b]{0.35\linewidth}
\begin{tikzpicture}
\node[loc] (ax) {$a$};
\node[loc,right=\ndist of ax] (ax1) {};
\node[loc,right=\ndist of ax1] (ax2) {$c$};
\node[right=\ndist of ax2] (ay) {$\nil$};
\node[loc,below=\ndist of ax] (ab) {$b$};
\node[loc,right=\ndist of ab] (anil) {};
\draw (ax) edge[->] (ax1);
\draw (ax1) edge[->] (ax2);
\draw (ax2) edge[->] (ay);
\draw (ab) edge[->] (anil);
\draw (anil) edge[->] (ax2);
\end{tikzpicture}
\end{minipage}\\
\end{tabular}
    \caption{Two models of
      $(\ls(a,\nil) * \true) \wedge (\ls(b,\nil) * \true)$ for a stack
      with domain $a,b$ and a stack with domain $a,b,c$.
      \vspace{-0.5cm}}
    \label{fig:ex:ssl-vs-wsl}
  \end{figure}

\begin{example}\label{ex:ssl-vs-wsl}
  Let $\phi := a \neq b * (\ls(a,\nil) * \true) \wedge (\ls(b,\nil) * \true)$.
  In Fig.~\ref{fig:ex:ssl-vs-wsl}, we show two models of $\phi$.
  On the left, we assume that $a,b$ are the only program variables, whereas on the right, we assume that there is a third program variable $c$.

  Note that the latter model, where the two lists overlap, is possible in SSL \emph{only} because the lists come together at the location labeled by $c$.
  If we removed variable $c$ from the stack, the model would no longer satisfy $\phi$ according to the strong semantics, because $\oursunion$ would no longer allow splitting the heap at that location.
  Conversely, the model would still satisfy $\phi$ with standard semantics.

  This is a feature rather than a bug of SSL:
  By demanding that the user of SSL specify aliasing explicitly---for example by using the specification $\ls(a,c)*\ls(b,c)*\ls(c,\nil) \wedge c \neq \nil$---we rule out unintended aliasing effects.
  \qedef{}
\end{example}

\paragraph*{Satisfiability and Semantic Consequence.}
We define the notions of satisfiability and semantic consequence parameterized by a finite set of variables $\vec{x} \subseteq \Var$.
For a formula $\phi$ with $\fvs{\phi} \subseteq \vec{x}$, we say that $\phi$ is \emph{satisfiable} w.r.t. $\vec{x}$ if there is a model $\SH$ with $\dom(\S) = \vec{x}$ such that $\SH \ourmodels \phi$.
We say that $\phi$ \emph{entails} $\psi$ w.r.t. $\vec{x}$, in signs $\phi \entailsx \psi$, if $\SH \ourmodels \phi$ then also $\SH \ourmodels \psi$ for all models $\SH$ with $\dom(\S) = \vec{x}$.

\subsection{Correspondence of Strong and Weak Semantics on Positive Formulas}\label{sec:correspondence}

We call an SL formula $\phi$ \emph{positive} if it does not contain $\neg$.
Note that, in particular, this implies that $\phi$ does \emph{not} contain the magic wand $\mw$ or the atom $\true$.

In models of positive formulas, all dangling locations are labeled by variables:

\begin{lemma}\label{lem:positive-dangling-labeled}
  Let $\phi$ be positive and $\SH \stdmodels \phi$.
  Then, $(\img(\H) \setminus \dom(\H)) \subseteq \img(\S)$.
\end{lemma}
\begin{AppendixOnly}
  \begin{appendixproof}
    \emph{(Lemma~\ref{lem:positive-dangling-labeled}.)}
    We prove the following stronger statement by structural induction on $\phi$:
    For every model $\SH \stdmodels \phi$ we have that
    \begin{enumerate}
      \item $(\img(\H) \setminus \dom(\H)) \subseteq \img(\S)$,
      \item every \emph{join point} is labelled by a variable, i.e., $\size{\H^{-1}(\loc)} \ge 2$ implies that $\loc \in \img(\S)$, and
      \item every \emph{source} is labelled by a variable, i.e., $\H(\loc) \neq \bot$ and $\size{\H^{-1}(\loc)} = 0$ imply that $\loc \in \img(\S)$.
    \end{enumerate}
    The proof is straightforward except for the $\sept$ case:
    Assume $\SH \stdmodels \phi_1 \sept \phi_2$.
    Then there is a $\H_0$ with $\SHpair{\S}{\H_0} \stdmodels \phi_1$ and $\SHpair{\S}{\H_0 \stdunion \H} \stdmodels \phi_2$.
    By induction assumption the claim holds for $\SHpair{\S}{\H_0}$ and $\SHpair{\S}{\H_0 \stdunion \H}$.
    We note that every join point of $\H$ is also a join point of $\H_0 \stdunion \H$ and hence labelled by a variable.
    We now verify that every pointer that is dangling in $\H$ is either also dangling in $\H_0 \stdunion \H$ or is a join point in $\H_0 \stdunion \H$ or is pointing to a source of $\H_0$;
    in all cases the target of the dangling pointer is labelled by a variable.
    Finally, a source of $\H$ is either also a source of $\H_0 \stdunion \H$ or is pointed to by a dangling pointer of $\H_0$;
    in both cases the source is labelled by a variable.
  \end{appendixproof}
\end{AppendixOnly}

As every location shared by heaps $\H_1$ and $\H_2$ in $\H_1 \stdunion \H_2$ is either dangling in $\H_1$ or in $\H_2$ (or in both), the operations $\stdunion$ and $\oursunion$ coincide on models of positive formulas:

\begin{lemma}\label{lem:positive-coinciding-unions}
  Let $\SHi{1}\stdmodels\phi_1$ and $\SHi{2}\stdmodels\phi_2$ for
  positive formulas $\phi_1,\phi_2$.
  Then $\H_1\stdunion\H_2\neq\bot$
  iff $\H_1 \oursunion\H_2\neq\bot$.
\end{lemma}
\begin{AppendixOnly}
  \begin{appendixproof}
    \emph{(Lemma~\ref{lem:positive-coinciding-unions}.)}
    If $\H_1\oursunion\H_2\neq\bot$, then
    $\H_1 \stdunion\H_2 \neq \bot$ by definition.

    Conversely, assume $\H_1\stdunion\H_2\neq\bot$. We need to show
    that $\locs(\H_1)\cap\locs(\H_2) \subseteq \img(\S)$.
    To this end, let $l \in \locs(\H_1)\cap\locs(\H_2)$. Then there
    exists an $i \in \set{1,2}$ such that
    $i \in \img(\H_i)\setminus\dom(\H_i)$---otherwise $l$ would be in
    $\dom(\H_1)\cap\dom(\H_2)$ and $\H_1\stdunion\H_2=\bot$.
    By Lemma~\ref{lem:positive-dangling-labeled}, we thus have
    $l \in \img(\S)$.
  \end{appendixproof}
\end{AppendixOnly}

Since the semantics coincide on atomic formulas by definition and on
$*$ by Lemma~\ref{lem:positive-dangling-labeled}, we can easily show
that they coincide on all positive formulas:

\begin{lemma}\label{lem:positive-same-semantics}
  Let $\phi$ be a positive formula and let $\SH$ be a model.
  Then,
  $\SH \stdmodels \phi$ iff $\SH \ourmodels \phi$.
\end{lemma}
\begin{AppendixOnly}
\begin{appendixproof}
  \emph{(Lemma~\ref{lem:positive-same-semantics}.)}
  We proceed by structural induction on $\phi$.
  If $\phi$ is atomic, there is nothing to show.
  For $\phi = \phi_1*\phi_2$ and $\phi = \phi_1\sept\phi_2$, the claim
  follows from the induction hypotheses and
  Lemma~\ref{lem:positive-coinciding-unions}.
  For $\phi=\phi_1\wedge\phi_2$ and $\phi=\phi_1\vee\phi_2$, the claim
  follows immediately from the induction hypotheses and the semantics
  of $\wedge$, $\vee$.
 
\end{appendixproof}
\end{AppendixOnly}
\noindent
Lemma~\ref{lem:positive-same-semantics} implies that the two semantics coincide on the popular \emph{symbolic-heap fragment} of separation logic.\footnote{Strictly speaking, Lemma~\ref{lem:positive-same-semantics} implies this only for the symbolic-heap fragment of the separation logic studied in this paper, i.e., with the list predicate but no other data structures.
The result can, however, be generalized to symbolic heaps with trees (see the dissertation of the first author~\cite{pagel2020decision}).
Symbolic heaps of bounded treewidth as proposed in~\cite{iosif2013tree} are an interesting direction for future work.}
Further, by negating Lemma~\ref{lem:positive-same-semantics},
we have that
$\set{\SH \mid \SH \stdmodels \phi} \neq
  \set{\SH \mid \SH \ourmodels \phi}$
implies that $\phi$ contains negation, either explicitly or in the form of a magic wand or $\true$.

We remark that formula $\phi$ in Example~\ref{ex:ssl-vs-wsl} only employs $\true$ but not $\neg, \mw$.
Hence, even if only $\true$ would be added to the positive fragment, Lemma~\ref{lem:positive-same-semantics} would no longer hold.
Likewise, Lemma~\ref{lem:positive-same-semantics} does not hold under intuitionistic semantics:
as the intuitionistic semantics of a predicate $p$ is equivalent to $p * \true$ under classic semantics, it is sufficient to consider $\phi := a \neq b * (\ls(a,\nil) \wedge (\ls(b,\nil))$.



\section{Deciding the SSL Satisfiability Problem}\label{sec:deciding}
The goal of this section is to develop a decision procedure for SSL:

\begin{theorem}\label{thm:sat:pspace-complete}
  Let $\phi \in \SL$ and let $\vec{x} \subseteq \Var$ be a finite set of variables with $\fvs{\phi} \subseteq \vec{x}$.
  It is decidable in $\PSpace$ (in $\size{\phi}$ and  $\size{\vec{x}}$) whether there exists a model $\SH$ with $\dom(\S) = \vec{x}$ and $\SH \ourmodels \phi$.
\end{theorem}
\noindent Our approach is based on abstracting stack--heap models by
\emph{abstract memory states} (AMS), which have two key properties, which together imply Theorem~\ref{thm:sat:pspace-complete}:
\begin{description}
\item[Refinement (Theorem~\ref{thm:ams-refinement}).]
  If $\SiHi{1}$ and $\SiHi{2}$ abstract to the same AMS, then they satisfy the same formulas.
  That is, the AMS abstraction \emph{refines} the  satisfaction relation of SSL.
\item[Computability (Theorem~\ref{thm:ssl:sat-decidable},  Lemmas~\ref{lem:sat:pspace-hard} and~\ref{lem:ssl:sat-pspace}).]
  For every formula $\phi$, we can compute (in $\PSpace$) the set of all AMSs of all models of $\phi$;
  then, $\phi$ is satisfiable if this set is nonempty.
\end{description}

The AMS abstraction is motivated by the following insights.
\begin{enumerate}
\item The operator $\oursunion$ induces a unique decomposition of the heap into at most $\size{\S}$ minimal \emph{chunks} of memory that cannot be further decomposed.
\item To decide whether $\SH \ourmodels \phi$ holds, it is sufficient to know for each chunk of $\SH$ a) which atomic formulas the chunk satisfies and b) which variables (if any) are allocated in the
  chunk.
\end{enumerate}

We proceed as follows.
In Sec~\ref{sec:memchunks}, we make precise the notion of memory chunks.
In Sec.~\ref{sec:ams}, we define \emph{abstract memory states} (AMS), an abstraction of models that retains for every chunk precisely the information from point (2)~above.
We will prove the \emph{refinement theorem} in \ref{sec:ssl-refinement}.
We will show in Sections~\ref{sec:ams-computation}--\ref{sec:algorithm} that we can compute the AMS of the models of a given formula $\phi$, which allows us to decide satisfiability and entailment problems for SSL.
Finally, we prove the $\PSpace$-completeness result in
Sec.~\ref{sec:complexity}.

\subsection{Memory Chunks}\label{sec:memchunks}
We will abstract a model $\SH$ by abstracting every \emph{chunk} of $\H$, which is a \emph{minimal} nonempty sub-heap of $\SH$ that can be split off of $\H$ according to the strong-separation semantics.

\begin{definition}[Sub-heap]
  \label{def:chunk}
  Let $\SH$ be a model.
  We say that $\H_1$ is a \emph{sub-heap} of $\H$,
  in signs $\H_1 \sqsubseteq \H$,
  if there is some heap $\H_2$ such that $\H = \H_1 \oursunion \H_2$.
  We collect all sub-heaps in the set $\subheaps\SH$.
   \qedef{}
\end{definition}
\noindent The following proposition is an immediate consequence of the above definition:
\begin{proposition}
  Let $\SH$ be a model.
  Then, $(\subheaps\SH,\sqsubseteq,\sqcup,\sqcap,\neg)$ is a Boolean algebra with greatest element $\H$ and smallest element $\emptyset$, where
  \begin{itemize}
    \item $(\S,\H_1) \sqcup (\S,\H_2) := (\S,\H_1 \cup \H_2)$,
    \item $(\S,\H_1) \sqcap (\S,\H_2) := (\S,\H_1 \cap \H_2)$, and
    \item $\neg (\S,\H_1) := (\S,\H_1')$, where $\H_1' \in \subheaps\SH$ is the unique sub-heap with $\H = \H_1 \oursunion \H_1'$.
  \end{itemize}
\end{proposition}

The fact that the sub-models form a Boolean algebra allows us to make the following definition\footnote{It is an interesting question for future work to relate the chunks considered in this paper to the atomic building blocks used in SL symbolic executions engines. Likewise, it would be interesting to build a symbolic execution engine based on the chunks resp. on the AMS abstraction proposed in this paper.}:

\begin{definition}[Chunk]
  \label{def:chunk}
  Let $\SH$ be a model.
  A \emph{chunk of $\SH$} is an atom of the Boolean algebra $(\subheaps\SH,\sqsubseteq,\sqcup,\sqcap,\neg)$.
  We collect all chunks of $\SH$ in the set $\chunks\SH$. \qedef{}
\end{definition}

Because every element of a Boolean algebra can be uniquely decomposed into atoms, we obtain that every heap can be fully decomposed into its chunks:

\begin{proposition}
\label{prop:chunk-decomp}
  Let $\SH$ be a model and let  $\chunks\SH=\set{\H_1,\ldots,\H_n}$ be its chunks.
  Then, $\H = \H_1 \oursunion \H_2 \oursunion \cdots \oursunion \H_n$.
\end{proposition}

\begin{example}\label{ex:chunks}
  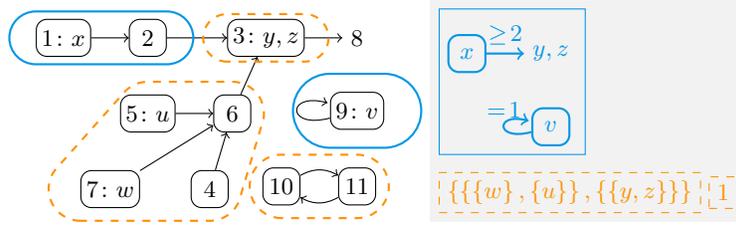
\begin{figure}[tb!]
    \centering
    \begin{tikzpicture}[node distance=5mm]
      \node[loc] (1) {$1\colon x$};
      \node[loc,right=of 1] (2) {$2$};
      \node[loc,right=8mm of 2] (3) {$3\colon y,z$};
      \node[loc,below=of 2] (11) {$5\colon u$};
      \node[loc,below=of 11,xshift=-5mm] (5) {$7\colon w$};
      \node[loc,right=of 11] (6) {$6$};
      \node[loc,below=of 6,xshift=-3mm] (4) {$4$};
      \node[right=of 3] (7) {$8$};
      \node[loc,below=of 7] (8) {$9\colon v$};
      \draw (1) edge[hedge] (2);
      \draw (2) edge[hedge] (3);
      \draw (4) edge[hedge] (6);
      \draw (5) edge[hedge] (6);
      \draw (6) edge[hedge] (3);
      \draw (3) edge[hedge] (7);
      \draw (11) edge[hedge] (6);
      \draw (8) edge[hedge,loop left] (8);

      \node[loc,below=of 8] (10) {$11$};
      \node[loc,left=of 10] (9) {$10$};
      \draw (9) edge[hedge,bend left] (10);
      \draw (10) edge[hedge,bend left] (9);

      \draw[azul1,thick]\convexpath{1.west,2.east}{10pt};
      \draw[laranja1,thick, dashed]\convexpath{5.west,11,6,4}{12pt};
      \draw[laranja1,thick, dashed]\convexpath{3.west,3.east}{9pt};
      \draw[laranja1,thick, dashed]\convexpath{9,10}{12pt};
      \draw[azul1,thick]\convexpath{8.west,8.east}{14pt};

      \node[right=of 8] {
        \tikzAMS{
          \begin{tikzpicture}
      \node[rgnode,right=2cm of 7,label={}] (r1) {$x$};
      \node[right=of r1,label={}] (r2) {$y,z$};
      \node[rgnode,below=of r2] (r3) {$v$};
      \draw (r1) edge[rgedge] node[above] {$\geqtwo$} (r2);
      \draw (r3) edge[rgedge,loop left] node[above] {$\eqone$} (r3);
    \end{tikzpicture}
  }{
      $\set{\set{\set{w},\set{u}}, \set{\set{y,z}}}$
    }{
      $1$
      }
      };
    \end{tikzpicture}

    \caption{Graphical representation of a model consisting of five chunks (left, see Ex.~\ref{ex:chunks}) and its induced AMS (right, see Ex.~\ref{ex:induced-ams}).
    \vspace{-0.5cm}}
    \label{fig:ex:chunks}
  \end{figure}
  Let $\S = \{x \mapsto 1, y \mapsto 3, u \mapsto 5, z \mapsto 3, w \mapsto 7, v \mapsto 9\}$ and $\H = \{1 \mapsto 2, 2\mapsto 3, 3 \mapsto 8, 4 \mapsto 6, 5 \mapsto 6, 6 \mapsto 3, 7 \mapsto 6, 9 \mapsto 9, 10\mapsto 11, 11\mapsto 10\}$.
  The model $\SH$ is illustrated in Fig.~\ref{fig:ex:chunks}.
  This time, we include the identities of the locations in the graphical representation;
  e.g., $3\colon y,z$ represents location $3$, $\S(y)=3$, $\S(z)=3$.
  The model consists of five chunks,
  $\H_1 := \{1 \mapsto 2, 2 \mapsto 3\}$,
  $\H_2 := \{9 \mapsto 9\}$,
  $\H_3 := \{4\mapsto 6, 5\mapsto 6, 6\mapsto 3, 7\mapsto 6\}$,
  $\H_4 := \{3 \mapsto 8\}$, and
  $\H_5 := \{10 \mapsto 11, 11 \mapsto 10\}$.
  \qedef{}
\end{example}

We distinguish two types of chunks:
those that satisfy SSL atoms and those that don't.

\begin{definition}[Positive and Negative chunk]
  Let $\H_c \subseteq \H$ be a chunk of $\SH$.
  $\H_c$ is a \emph{positive chunk} if there exists an atomic formula $\tau$ such that  $\SHpair{\S}{\H_c}\ourmodels \tau$.
  Otherwise, $\H_c$ is a \emph{negative chunk}.
  We collect the respective chunks in $\goodchunks\SH$ and $\badchunks\SH$.
\end{definition}

\begin{example}\label{ex:chunks:path}
  Recall the chunks $\H_1$ through $\H_5$ from Ex.~\ref{ex:chunks}.
  $\H_1$ and $\H_2$ are positive chunks (blue in
  Fig.~\ref{fig:ex:chunks}), $\H_3$ to $\H_5$ are negative chunks (orange).
  \qedef{}
\end{example}


Negative chunks fall into three (not mutually-exclusive) categories:
\begin{description}
\item[Garbage.] Chunks with locations that are inaccessible via stack variables.
\item[Unlabeled dangling pointers.] Chunks with an unlabeled sink, i.e., a dangling location that is not in $\img(\S)$ and thus \emph{cannot} be ``made non-dangling'' via composition using $\oursunion$.
\item[Overlaid list segments.] Overlaid list segments that cannot be separated via $\oursunion$ because they are joined at locations that are not in $\img(\S)$.
\end{description}
\begin{example}[Negative chunks]
  The chunk $\H_3$ from Example~\ref{ex:chunks} contains garbage, namely the location $4$ that cannot be reached via stack variables, \emph{and} two overlaid list segments (from $5$ to $3$ and $7$ to $3$).
  The chunk $\H_4$ has an unlabeled dangling pointer.
  The chunk $\H_5$ contains only garbage.
\end{example}

\subsection{Abstract Memory States}\label{sec:ams}
\newcommand{\AMquadruple}{\tuple{\V,\E,\Rset,\garb}}
\newcommand{\AMquadruplei}[1]{\tuple{\V_{#1},\E_{#1},\Rset_{#1},\garb_{#1}}}
\newcommand{\eqclasses}[1]{\mathsf{cls}_{=}(#1)}
\newcommand{\eqclass}[2]{{[#2]}^{#1}_{=}}
\newcommand{\phiabstkw}{\alpha}
\newcommand{\phiabsts}[2]{\phiabstkw_{#1}(#2)}
\newcommand{\phiabstx}[2]{\phiabstkw_{#1}(#2)}

In \emph{abstract memory states} (AMSs), we retain for every chunk enough information to (1) determine which atomic formulas the chunk satisfies, and (2) keep track of which variables are allocated within each chunk.

\begin{definition}\label{def:ams}
  A quadruple $\A = \AMquadruple$ is an \emph{abstract memory state}, if
  \begin{enumerate}
  \item $\V$ is a \emph{partition} of some finite set of variables, i.e., $\V = \{\vec{v}_1, \ldots,\vec{v}_n\}$ for some non-empty disjoint finite sets $\vec{v}_i \subseteq \Var$,
  \item $\E \colon \V \pfun \V \times \theListInts$ is a partial function such that there is
      no $\vec{v} \in \dom(\E)$ with $\nil \in \vec{v}$\footnote{The edges of an AMS represent either a single pointer (case ``$\eqone$'') or a list segment of at least length two (case ``$\geqtwo$'').},
  \item $\Rset$ consists of disjoint subsets of $V$ such that every $R \in \Rset$ is disjoint from $\dom(\E)$ and there is no $\vec{v} \in R$ with $\nil \in \vec{v}$,
  \item $\garb$ is a natural number, i.e., $\garb \in \N$.
  \end{enumerate}
 We call $\V$ the \emph{nodes}, $\E$ the \emph{edges}, $\Rset$ the \emph{negative-allocation constraint} and $\garb$ the \emph{garbage-chunk count} of $\A$.
 We call the AMS $\A=\AMquadruple$ \emph{garbage-free} if $\Rset=\emptyset$ and $\garb=\emptyset$.

 We collect the set of all AMSs in $\AMS$.
 The \emph{size} of $\A$ is given by
 $\size{\A}:=\size{\V} + \garb$.
 Finally, the \emph{allocated variables} of an AMS are given by
 $\alloc{\A} := \dom(\E) \cup \bigcup \Rset$.
  \qedef{}
\end{definition}

Every model induces an AMS, defined in terms of the following auxiliary definitions.
The equivalence class of variable $x$ w.r.t.~stack $\S$ is
$\eqclass{\S}{x} := \set{y \mid \S(y)=\S(x)}$;
the set of all equivalence classes of $\S$ is
$\eqclasses\S := \set{\eqclass{\S}{x} \mid x \in \dom(\S)}$.
We now define the edges induced by a model $\SH$:
For every equivalence class $\eqclass{\S}{x} \in \eqclasses\S$, we set
\[
    \iedges\SH(\eqclass{\S}{x}) :=
    \begin{cases}
      \tuple{\eqclass{\S}{y}, \eqone} & \text{there are } y \in \dom(\S) \text{ and } \H_c \in \goodchunks\SH       \\
      &  \text{ with } (\S,\H_c) \ourmodels x \pto y \\
      \tuple{\eqclass{\S}{y}, \geqtwo} &  \text{there are } y \in \dom(\S) \text{ and } \H_c \in \goodchunks\SH
      \\
      &  \text{ with } (\S,\H_c) \ourmodels \ls(x,y) \wedge \neg x \pto y \\
     \bot, & \text{otherwise}.
    \end{cases}
\]
Finally, we denote the sets of variables allocated in negative chunks by
\[
\mfalloc\SH := \{ \{\eqclass{\S}{x} \mid \S(x) \in \dom(\H_c) \} \mid \H_c \in \badchunks\SH \} \setminus \{\emptyset\},
\]
where (equivalence classes of) variables that are allocated in the same negative chunk are grouped together in a set.

Now we are ready to define the \emph{induced AMS} of a model.
\begin{definition}
  Let $\SH$ be a model.
  Let $\V := \eqclasses\S$,
  $\E := \iedges\SH$,
  $\Rset :=\mfalloc\SH$ and
  $\garb := \size{\badchunks\SH}-\size{\mfalloc\SH}$.
  Then $\mabst\SH := \AMquadruple$ is the \emph{induced AMS} of $\SH$. \qedef
\end{definition}
\begin{example}\label{ex:induced-ams}
  The induced AMS of the model $\SH$ from Ex.~\ref{ex:chunks} is illustrated on the right-hand side of Fig.~\ref{fig:ex:chunks}.
  The blue box depicts the graph $(\V,\E)$ induced by the positive chunks $\H_1,\H_2$;
  the negative chunks that allocate variables are abstracted to the set $\Rset = \set{\set{\set{w},\set{u}}, \set{\set{y,z}}}$
  (note that the variables $w$ and $u$ are allocated in the chunk $\H_3$ and the aliasing variables $y,z$ are allocated in $\H_4$);
  and the garbage-chunk count is $1$, because $\H_5$ is the only negative chunk that does not allocate stack variables. \qedef
\end{example}
Observe that the induced AMS is indeed an AMS:
\begin{proposition}
  Let $\SH$ be a model.
  Then $\mabst\SH \in \AMS$.
\end{proposition}
The reverse also holds:
Every AMS is the induced AMS of at least one model;
in fact, even of a model of linear size.
\begin{lemma}[Realizability of AMS]\label{lem:ams:poly-models}
  Let $\A = \AMquadruple$ be an AMS.
  There exists a model $\SH$ with $\mabst\SH=\A$ whose size is linear in the size of $\A$.
\end{lemma}
\begin{AppendixOnly}
\begin{appendixproof}
  \emph{(Lemma~\ref{lem:ams:poly-models}.)}
  \newcommand{\vfun}{t}
  For simplicity, we assume $\Loc = \N$;
  this allows us to add locations.

  Let $n := \size{\V}$.
  We fix some injective function $\vfun\colon\V \to \set{1,\ldots,n}$ from nodes to natural numbers.
  We set $\S := \bigcup_{x \in v, v \in \V} \set{x \mapsto \vfun(v)}$ and define $\H$ as the (disjoint) union of
  \begin{itemize}
  \item $\bigcup_{\E(v)= \tuple{v',\eqone} } \set{ \vfun(v) \mapsto \vfun(v') }$
  \item $\bigcup_{\E(v)=\tuple{v',\geqtwo} } \set{ \vfun(v) \mapsto n + \vfun(v),\allowbreak n +
      \vfun(v)\allowbreak \mapsto \vfun(v') }$
  \item $\bigcup_{v \in \vec{r},\vec{r} \in \Rset}
  \set{\vfun(v) \mapsto 2n + \vfun(\max(\vec{r}))}$
  \item $\bigcup_{l \in \set{3n+1, \ldots, 3n+\garb}} \set{l \mapsto l}$
  \end{itemize}
It is easy to verify that $\mabst\SH=\A$ and that  $\size{\H} \in \bigO(\size{\A})$.
\end{appendixproof}
\end{AppendixOnly}
\begin{SubmissionOnly}
  We can construct such a model for $\A=\AMquadruple$ as
  follows.
  Let $\S$ be an arbitrary stack with  $\eqclasses{\S}=\V$.
  For every edge in $\E$, add one pointer to $\H$ if the edge is labeled by $\eqone$ and add two pointers to $\H$ if the edge is labeled by $\geqtwo$;
  for every set $\set{x_1,\ldots,x_n}$ in the malformed-allocation constraint, let $l$ be a fresh location and add at most $n$ pointers, one from each
  location $\S(x_i)$ to $l$;
  and for a garbage-chunk count of $k$, let $l_1,\ldots,l_{2k}$ be $2k$ pairwise different, fresh locations and add pointers $l_i \pto l_{k+i}$ to $\H$.
\end{SubmissionOnly}

The following lemma demonstrates that we only need the $\Rset$ and $\garb$ components in order to be able to deal with negation and/or the magic wand:

\begin{lemma}[Models of Positive Formulas Abstract to Garbage-free AMS]
\label{garbage-free-AMS}
  Let $(\S,\H)$ be a model.
  If $(\S,\H) \models \phi$ for a positive formula $\phi$, then $\mabst{\SH}$ is garbage-free.
\end{lemma}
\begin{appendixproof}
  \emph{(Lemma~\ref{garbage-free-AMS}.)}
  The lemma can be proved by a straight-forward induction on $\phi$, using that every heap fully decomposes into its chunks.
\end{appendixproof}

We abstract SL formulas by the set of AMS of their models:

\begin{definition}
  \label{def:sl:abstraction}
  Let $\S$ be a stack.
  The \emph{$\SL$ abstraction} w.r.t.~$\S$,
  $\phiabstkw_\S \colon \SL \to 2^{\AMS}$, is given by
  \begin{align*}
    &\phiabsts{\S}{\phi} := \{\mabst\SH \mid \H \in \HH,
    \text{ and } \SH
      \ourmodels \phi\}.\tag*{$\triangle$}
  \end{align*}
\end{definition}

Because AMSs do not retain any information about heap locations, just about aliasing, abstractions do not differ for stacks with the same equivalence classes:
\begin{lemma}\label{lem:same-eqclasses-same-mabst}
  Let $\S,\S'$ be stacks with $\eqclasses{\S}=\eqclasses{\S'}$.
  Then $\phiabsts{\S}{\phi}=\phiabsts{\S'}{\phi}$ for all formulas $\phi$.
\end{lemma}
\begin{AppendixOnly}
  \begin{appendixproof}
    \emph{(Lemma~\ref{lem:same-eqclasses-same-mabst}.)}
    Let $\A \in \phiabsts{\S}{\phi}$.
    Then there exists a heap $\H$ such that $\mabst\SH=\A$ and $\SH \ourmodels \phi$.
    Let $\H'$ be such that $\SH \Iso \SHpprime$.
    By Lemma~\ref{lem:iso-same-sat},
    $\SHpprime \ourmodels \phi$.
    Moreover, $\mabst\SHpprime=\A$.
    Consequently, $\A \in \phiabsts{\S'}{\phi}$.
    The other direction is proved analogously.
  \end{appendixproof}
\end{AppendixOnly}


\subsection{The Refinement Theorem for SSL}\label{sec:ssl-refinement}

The main goal of this section is to show the following
\emph{refinement theorem}:

\begin{theorem}[Refinement Theorem]
  \label{thm:ams-refinement}
  Let $\phi$ be a formula and let $\SHi{1}$, $\SHi{2}$ be models with $\mabst{\SHi{1}} = \mabst{\SHi{2}}$.
  Then $\SHi{1}\ourmodels\phi$ iff $\SHi{2}\ourmodels\phi$.
\end{theorem}

We will prove this theorem step by step, characterizing the AMS abstraction of all atomic formulas and of the composed models before proving the refinement theorem.
In the remainder of this section, we fix some model $\SH$.

\paragraph*{Abstract Memory States of Atomic Formulas}

The empty-heap predicate $\emp$ is only satisfied by the empty heap, i.e., by a heap that consists of zero chunks:

\begin{lemma}\label{lem:phiabst:emp}
  $\SH \models \emp$ iff $\mabst{\SH} = \tuple{\eqclasses\S, \emptyset, \emptyset, 0}$
\end{lemma}
\begin{appendixproof}
  \emph{(Lemma~\ref{lem:phiabst:emp}.)}
  $\SH\ourmodels \emp$ iff $\H = \emptyset$ iff $\chunks{\SH}=\emptyset$ iff
  $\mabst{\SH}=\tuple{\eqclasses\S, \emptyset, \emptyset, 0}$.
\end{appendixproof}

\begin{lemma}\label{lem:phiabst:eq}
  \begin{enumerate}
  \item $\SH \models x=y$ iff
    $\mabst{\SH}=\tuple{\eqclasses\S, \emptyset, \emptyset, 0}$
    and $\eqclass{\S}{x}=\eqclass{\S}{y}$.
  \item $\SH \models x \neq y$ iff
    $\mabst{\SH}=\tuple{\eqclasses\S, \emptyset, \emptyset, 0}$
    and $\eqclass{\S}{x}\neq\eqclass{\S}{y}$.
  \end{enumerate}
\end{lemma}
\begin{appendixproof}
  \emph{(Lemma~\ref{lem:phiabst:eq}.)}
  We only show the first claim, as the proof of the second claim is completely analogous.
  $\SH \models x=y$ iff ($\S(x)=\S(y)$ and $\H = \emptyset$)
  iff ($\eqclass{\S}{x}=\eqclass{\S}{y}$ and $\SH\models \emp$)
  iff, by Lemma~\ref{lem:phiabst:emp},
  ($\eqclass{\S}{x}=\eqclass{\S}{y}$ and
  $\mabst{\SH}=\tuple{\eqclasses\S, \emptyset, \emptyset, 0}$).
\end{appendixproof}

Models of points-to assertions consist of a single positive chunk of size $1$:

\begin{lemma}\label{lem:phiabst:pto}
Let $\E = \{ \eqclass{\S}{x} \mapsto
      \tuple{\eqclass{\S}{y}, \eqone}\}$.
$\SH \models x \pto y$ iff
$\mabst{\SH} = \langle \eqclasses\S, \E, \emptyset, 0 \rangle$.
\end{lemma}
\begin{appendixproof}
  \emph{(Lemma~\ref{lem:phiabst:pto}.)}
  If $\SH\ourmodels x \pto y$ then $\H = \set{\S(x) \mapsto \S(y)}$.
  In particular, it then holds that $\H$ is a positive chunk.
  Consequently, $\iedges\SH = \E$.
  It follows that $\mabst{\SH}=\langle \eqclasses\S, \E,\allowbreak \emptyset, 0 \rangle$.

  Conversely, assume
  $\mabst{\SH}=\langle \eqclasses\S, \E,\allowbreak \emptyset, 0 \rangle$.
  Then, $\SH$ consists of a single positive chunk and no negative chunks.
  Further, by the definition of $\iedges\SH$ we have that this single positive chunk satisfies $\SH \models x \pto y$.
\end{appendixproof}

Intuitively, the list segment $\ls(x,y)$ is satisfied by models $\SH$ that consist of zero or more positive chunks, corresponding to a (possibly empty) list from some equivalence class $\eqclass{\S}{x}$ to $\eqclass{\S}{y}$ via (zero or more) intermediate equivalence classes $\eqclass{\S}{x_1},\ldots,\eqclass{\S}{x_n}$.
We will use this intuition to define abstract lists;
this notion allows us to characterize the AMSs arising from abstracting lists.
\begin{definition}
  Let $\A=\AMquadruple\in\AMS$, $\S$ be a stack and $x,y \in \Var$.
  We say $A$ is an \emph{abstract list} w.r.t. $x$ and $y$, in signs $\A \in \AbstLists{x}{y}$, iff
  \begin{enumerate}
  \item $V = \eqclasses\S$,
  \item $\Rset=\emptyset$ and $\garb=0$, and
  \item we can pick nodes $\vec{v}_1,\ldots, \vec{v}_n \in \V$ and labels $\iota_1,\ldots,\iota_{n-1} \in \theListInts$ such that
      $x \in \vec{v}_1$,
      $y \in \vec{v}_n$ and
      $\E = \{\vec{v}_i \mapsto \tuple{\vec{v}_{i+1},\iota_i} \mid
      1 \le i < n \}$. \qedef
  \end{enumerate}
\end{definition}
\begin{lemma}\label{lem:phiabst:ls}
  $\SH \models \ls(x,y)$ iff $\mabst{\SH} \in \AbstLists{x}{y}$.
\end{lemma}
\begin{AppendixOnly}
  \begin{appendixproof}
    \emph{(Lemma~\ref{lem:phiabst:ls}.)}
    Assume $\SH \models \ls(x,y)$.
    By the semantics, there exist locations $\loc_0,\ldots,\loc_n$,
    $n \geq 1$, with $\S(x)=\loc_0$, $\S(y)=\loc_n$ and $\H = \{\loc_0 \mapsto \loc_1,\ldots,\loc_{n-1} \mapsto \loc_n \}$.
    Let $j_1,\ldots,j_k$ those indices among $1,\ldots,n$ with  $\loc_{j_i} \in \img(\S)$.
    (In particular, $j_1=1$ and $j_k=n$.)
    Then for each $j_i$, the restriction of $\H$ to
    $\loc_{j_i},\loc_{j_i+1},\ldots,\loc_{j_{i+1}-1}$ is a positive chunk that either satisfies a points-to assertion or a list-segment predicate.
    Hence,
    $\iedges\SH(\S^{-1}(\loc_{j_i})) =  \tuple{\S^{-1}(\loc_{j_{i+1}}),\iota_i}$ for all $1 \le i < k$, for some $\iota_i \in \theListInts$.
    Thus, $\mabst{\SH} \in \AbstLists{x}{y}$.

    Assume $\mabst{\SH} \in \AbstLists{x}{y}$.
    Then, there are equivalence classes $\eqclass{\S}{x_1},\ldots,\eqclass{\S}{x_n} \in \eqclasses\S$ and labels $\iota_1,\ldots,\iota_{n-1} \in \theListInts$ such that $x \in \eqclass{\S}{x_1}$, $y \in \eqclass{\S}{x_n}$ and $\iedges\SH = \{\eqclass{\S}{x_i} \mapsto \tuple{\eqclass{\S}{x_{i+1}},\iota_i} \mid 1 \le i < n \}$.
    By the definition of $\iedges\SH$, we have that there are positive chunks $\H_i$ of $\H$ such that $\SHi{i} \ourmodels x_i \pto x_{i+1}$ or $\SHi{i} \ourmodels \ls(x_i,x_{i+1})$.
    In particular, we have $\H_i = \set{\loc_{i,1}
                            \mapsto \loc_{i,2}, \ldots, \loc_{i,k_i-1}\mapsto
                            \loc_{i,k_i}}, \S(x_i)=\loc_{i,1} \text{
                            and } \S(x_{i+1})=\loc_{i,k_i}$ for some locations $\loc_{i,j}$
    Because $\H$ does not have negative chunks, we get that $\H$ fully decomposes into its positive chunks.
    Hence, the locations $\loc_{i,j}$ witness that $\SH \models \ls(x,y)$.
  \end{appendixproof}
\end{AppendixOnly}


\paragraph*{Abstract Memory States of Models composed by the Union Operator}

Our next goal is to lift the union operator $\oursunion$ to the abstract domain $\AMS$.
We will define an operator $\Compose$ with the following property:
\begin{align*}
\text{if } \H_1\oursunion\H_2\neq \bot \text{~~then~~} \mabst{(\S,\H_1 \oursunion \H_2)} = \mabst{\SHi{1}\Compose \SHi{2}}.
\end{align*}

AMS composition is a partial operation defined only on
\emph{compatible} AMS.
Compatibility enforces (1) that the AMSs were obtained for equivalent stacks (i.e., for stacks $\S,\S'$ with
$\eqclasses\S=\eqclasses{\S'}$), and (2) that there is no double allocation.

\begin{definition}[Compatibility of AMSs]
  AMSs $\A_1=\AMquadruplei{1}$ and $\A_2=\AMquadruplei{2}$ are \emph{compatible} iff (1) $\V_1=\V_2$ and (2)
  $\alloc{\A_1}\allowbreak\cap\alloc{\A_2}=\emptyset$.
\end{definition}

Note that if $\H_1 \oursunion \H_2$ is defined, then
$\mabst{\SHi{1}}$ and $\mabst{\SHi{2}}$ are compatible. The converse is not true, because $\mabst{\SHi{1}}$  and $\mabst{\SHi{2}}$ may be compatible even if $\dom(\H_1)\cap\dom(\H_2)\neq\emptyset$.

AMS composition is defined in a point-wise manner on compatible AMSs and undefined otherwise.

\begin{definition}[AMS composition]\label{def:ams:compose}
  Let $\A_1=\AMquadruplei{1}$ and $\A_2=\AMquadruplei{2}$ be two AMS.
  The \emph{composition} of $\A_1,\A_2$ is then given by
  \[
    \A_1 \Compose \A_2 := \begin{cases}
      \tuple{\V_1,\E_1\cup\E_2,\Rset_1\cup\Rset_2,
          \garb_1+\garb_2}, & \text{if } \A_1,\A_2 \text{ compatible}\\
        \bot, &\text{otherwise.}
    \end{cases}
  \]
\end{definition}

\begin{lemma}\label{lem:ams:partiality-lifted}
  Let $\S$ be a stack and let $\H_1,\H_2$ be heaps.
  If $\H_1\oursunion\H_2\neq\bot$ then $\mabst{\SHi{1}}\Compose\mabst{\SHi{2}}\neq\bot$.
\end{lemma}
  \begin{appendixproof}
    \emph{(Lemma~\ref{lem:ams:partiality-lifted}.)}
    Since the same stack $\S$ underlies both abstractions, we have $\V_1=\V_2$.
    Furthermore, $\dom(\H_1)\cap\dom(\H_2)=\emptyset$ implies that  $\alloc{\A_1}\cap\alloc{\A_2}=\emptyset$.
  \end{appendixproof}

We next show that $\mabst{(\S,\H_1\oursunion\H_2)} =
\mabst{\SHi{1}}\Compose\mabst{\SHi{2}}$ whenever $\H_1\oursunion\H_2$ is defined:

\begin{lemma}[Homomorphism of
  composition]\label{lem:compose:model-composition}
  Let $\SHi{1},\SHi{2}$ be models with $\H_1\oursunion\H_2\neq \bot$.
  Then,
  $\mabst{(\S,\H_1\oursunion\H_2)}=\mabst{\SHi{1}}\Compose\mabst{\SHi{2}}$.
\end{lemma}
\begin{appendixproof}
  \emph{(Lemma~\ref{lem:compose:model-composition}.)}
  The result follows easily from the observation that
  \[
    \chunks{\SHpair{\S}{\H_1\oursunion\H_2}} = \chunks{\SHi{1}} \cup \chunks{\SHi{2}},
  \]
  which, in turn, is an immediate consequence of Proposition~\ref{prop:chunk-decomp}.
\end{appendixproof}

To show the refinement theorem, we need one additional property of AMS composition.
If an AMS $\A$ of a model $\SH$ can be decomposed into
two smaller AMS $\A=\A_1\Compose\A_2$, it is also possible to decompose the heap $\H$ into smaller heaps $\H_1,\H_2$ with $\mabst{\SHi{i}}=\A_i$:

\begin{lemma}[Decomposability of AMS]
\label{lem:decompose}
  Let $\mabst{\SH}=\A_1\Compose\A_2$.
  There exist $\H_1,\H_2$ with $\H=\H_1\oursunion\H_2$, $\mabst{\SHi{1}}=\A_1$ and $\mabst{\SHi{2}}=\A_2$.
\end{lemma}
\begin{appendixproof}
  \emph{(Lemma~\ref{lem:decompose}.)}
  It can be verified from the definition of AMS and the definition of the composition operator $\Compose$ that the following property holds:
  Let $\H_c \in \chunks\SH$ be a chunk.
  Then,  either there exists an $\A_1'$
  such that $\A_1 = \mabst{(\S,\H_c)} \Compose \A_1'$ or there exists an $\A_2'$ such that $\A_2 = \mabst{(\S,\H_c)} \Compose \A_2'$.

  The claim can then by proven an induction of the number of chunks $\size{\chunks\SH}$.
\end{appendixproof}

These results suffice to prove the Refinement Theorem stated at the beginning of this section (see the extended version~\cite{journals/corr/abs-2001-06235} for a proof).

\begin{appendixproof}
  \emph{(Theorem~\ref{thm:ams-refinement}.)}
  \newcommand{\SHtwoprime}{\SHpair{\S}{\H_2'}}
  Let $\A = \AMquadruple$ be the AMS with
  $\mabst{\SHi{1}} = \A = \mabst{\SHi{2}}$.
  We proceed by induction on the structure of
  $\phi$.
  We only prove that $\SHi{1} \ourmodels \phi$ implies that $\SHi{2}\ourmodels \phi$, as the other direction is completely analogous.

  Assume that the claim holds for all subformulas of $\phi$ and assume that $\SHi{1} \ourmodels \phi$.
  \begin{description}
  \item[Case $\emp$, $x = y$, $x \neq y$, $x \pto y$, $\ls(x,y)$.] Immediate consequence of Lemmas~\ref{lem:phiabst:emp},~\ref{lem:phiabst:eq},~\ref{lem:phiabst:pto} and~\ref{lem:phiabst:ls}.
  \item[Case $\phi_1 * \phi_2$.]
    By the semantics of $*$, there exist $\Hoo,\Hot$
    with $\H_1 = \Hoo \oursunion \Hot$,
    $\SHpair{\S}{\Hoo}\ourmodels \phi_1$, and
    $\SHpair{\S}{\Hot}\ourmodels \phi_2$.
    Let $\A_1 := \mabst{(\S,\Hoo)}$ and
    $\A_2 := \mabst{(\S,\Hot)}$.
    By Lemma~\ref{lem:compose:model-composition},
    $\mabst{\SHi{1}} = \A_1 \Compose \A_2 = \mabst{\SHi{2}}$.
    We can thus apply Lemma~\ref{lem:decompose} to $\mabst{\SHi{2}}$, $\A_1$, and $\A_2$ to obtain heaps $\Hto,\Htt$ with $\H_2=\Hto\oursunion\Htt$, $\mabst{(\S,\Hto)}=\A_1$ and $\mabst{(\S,\Htt)}=\A_2$.
    We can now apply the induction hypotheses for $1 \leq i \leq 2$, $\phi_i$, $\H_{1,i}$ and $\H_{2,i}$, and obtain that $\SHpair{\S}{\H_{2,i}}\ourmodels\phi_i$.
    By the semantics of $*$, we then have
    $\SHi{2}=\SHpair{\S}{\Hto \oursunion \Htt}\ourmodels
    \phi_1 * \phi_2$.
  \item[Case $\phi_1 \sept \phi_2$.]
    Since $\SHi{1}\ourmodels\phi_1\sept\phi_2$, there exists a heap $\H_0$ with $\SHpair{\S}{\H_0} \ourmodels \phi_1$ and $\SHpair{\S}{\H_1 \oursunion \H_0} \ourmodels \phi_2$.
    We can assume w.l.o.g.~that $\H_2\oursunion \H_0 \neq \bot$---if this is not the case, simply replace $\H_0$ with a heap $\H_0'$ with $\SHpair{\S}{\H_0}\Iso\SHpair{\S}{\H_0'}$,
    $\H_1\oursunion \H_0' \neq \bot$ and $\H_2\oursunion \H_0' \neq \bot$;
    then, $\SHpair{\S}{\H_1 \oursunion \H_0'} \ourmodels \phi_2$ by Lemma~\ref{lem:iso-same-sat}.
    We have that $\mabst{(\S,\H_1 \oursunion \H_0)} =
    \mabst{\SHi{1}} \Compose \mabst{(\S,\H_0)} = \mabst{\SHi{2}} \Compose \mabst{(\S,\H_0)} = \mabst{(\S,\H_2 \oursunion \H_0)}$ (by assumption and Lemma~\ref{lem:compose:model-composition}).
    It therefore follows from the induction hypothesis for $\phi_2$, $\SHpair{\S}{\H_1 \oursunion \H_0}$, and $\SHpair{\S}{\H_2 \oursunion \H_0}$ that $\SHpair{\S}{\H_2 \oursunion \H_0}\ourmodels \phi_2$.
    Thus, $\SHi{2}\ourmodels\phi_1\sept\phi_2$.

  \item[Case $\phi_1 \wedge \phi_2$, $\phi_1 \vee \phi_2$.]
    By the semantics of $\wedge$ resp. $\vee$,
    we have $\SHi{1}\ourmodels \phi_1$ and/or
    $\SHi{1}\ourmodels \phi_2$.
    We apply the induction hypotheses for $\phi_1$ and
    $\phi_2$ to obtain $\SHi{2}\ourmodels \phi_1$ and/or
    $\SHi{2}\ourmodels \phi_2$.
    By the semantics of $\wedge$ resp. $\vee$, we then have $\SHi{2}\ourmodels\phi_1\wedge\phi_2$ resp. $\SHi{2}\ourmodels\phi_1\vee\phi_2$.

  \item[Case $\neg \phi_1$.]
    By the semantics of $\neg$, we have $\SHi{1}\not\ourmodels \phi_1$.
    By the induction hypothesis for $\phi_1$ we then obtain $\SHi{2}\not\ourmodels \phi_1$.
    By the semantics of $\neg$, we have $\SHi{2}\ourmodels\neg\phi_1$.
    \qedhere
  \end{description}
\end{appendixproof}

\begin{corollary}\label{cor:same-abst--same-sat}
  Let $\SH$ be a model and $\phi$ be a formula.
  $\SH \ourmodels \phi$ iff
  $\mabst{\SH} \in \phiabsts{\S}{\phi}$.
\end{corollary}
\begin{appendixproof}
  \emph{(Corollary~\ref{cor:same-abst--same-sat}.)}
  Assume $\A := \mabst{\SH} \in \phiabsts{\S}{\phi}$.
  By definition of $\phiabstkw_\S$ there is a model $\SHprime$ with $\SHprime \ourmodels \phi$ and $\mabst{(\S,\H')} = \A$.
  By applying Theorem~\ref{thm:ams-refinement} to $\phi$, $\SH$ and $\SHprime$, we then get that $\SH \ourmodels \phi$. 
\end{appendixproof}

\subsection{Recursive Equations for Abstract Memory States}\label{sec:ams-computation}

In this section, we derive recursive equations that reduce the set of AMS $\phiabsts{\S}{\phi}$ for arbitrary compound formulas to the set of AMS of the constituent formulas of $\phi$.
In the next sections, we will show that we can actually evaluate these equations, thus obtaining an algorithm for computing the abstraction of arbitrary formulas.

\begin{SubmissionOnly}
  For example, $\SH \ourmodels \neg \phi_1$ iff it is not the case
  that $\SH\ourmodels \phi_1$ iff it is not the case that
  $\mabst\SH \in \phiabsts{\S}{\phi_1}$ iff it is the case that
  $\mabst\SH \in \set{\mabst\SH \mid \H \in \HH} \setminus \phiabsts{\S}{\phi_1}$.
\end{SubmissionOnly}

\begin{lemma}\label{lem:phiabst:wedge}
  $\phiabsts{\S}{\phi_1\wedge\phi_2} = \phiabsts{\S}{\phi_1}\cap\phiabsts{\S}{\phi_2}$.
\end{lemma}
\begin{AppendixOnly}
  \begin{appendixproof}
  \emph{(Lemma~\ref{lem:phiabst:wedge}.)}
  Let $\SH$ be a model.
  $\SH \ourmodels \phi_1 \wedge \phi_2$ iff
  $\SH\ourmodels \phi_1$ and $\SH\ourmodels \phi_2$ iff
  $\mabst\SH \in \phiabsts{\S}{\phi_1}$ and $\mabst\SH \in
  \phiabsts{\S}{\phi_2}$ iff
  $\mabst\SH \in \phiabsts{\S}{\phi_1} \cap \phiabsts{\S}{\phi_2}$.
  \end{appendixproof}
\end{AppendixOnly}

\begin{lemma}\label{lem:phiabst:vee}
  $\phiabsts{\S}{\phi_1\vee\phi_2} = \phiabsts{\S}{\phi_1}\cup\phiabsts{\S}{\phi_2}$.
\end{lemma}
\begin{AppendixOnly}
  \begin{appendixproof}
    \emph{(Lemma~\ref{lem:phiabst:vee}.)}
    Let $\SH$ be a model.
    $\SH \ourmodels \phi_1 \wedge \phi_2$ iff ($\SH\ourmodels \phi_1$
    or $\SH\ourmodels \phi_2$) iff
    ($\mabst\SH \in \phiabsts{\S}{\phi_1}$ or
    $\mabst\SH \in \phiabsts{\S}{\phi_2}$) iff
    $\mabst\SH \in \phiabsts{\S}{\phi_1} \cup \phiabsts{\S}{\phi_2}$.
  \end{appendixproof}
\end{AppendixOnly}

\begin{lemma}\label{lem:phiabst:neg}
  $\phiabsts{\S}{\neg\phi_1} = \set{\mabst\SH \mid \H \in \HH} \setminus
  \phiabsts{\S}{\phi_1}$.
\end{lemma}
\begin{AppendixOnly}
  \begin{appendixproof}
  \emph{Lemma~\ref{lem:phiabst:neg}.}
  Let $\SH$ be a model.
  $\SH \ourmodels \neg \phi_1$ iff it is not the case
  that $\SH\ourmodels \phi_1$ iff it is not the case that
  $\mabst\SH \in \phiabsts{\S}{\phi_1}$ iff it is the case that
  $\mabst\SH \in \set{\mabst\SH \mid \H \in \HH} \setminus \phiabsts{\S}{\phi_1}$.
  \end{appendixproof}
\end{AppendixOnly}


\paragraph*{The Separating Conjunction}

In Section~\ref{sec:ssl-refinement}, we defined the composition operation, $\Compose$, on pairs of AMS.
We now lift this operation to sets of AMS $\Aset_1, \Aset_2$:
\[\Aset_1 \Compose \Aset_2 := \set{\A_1 \Compose \A_2 \mid \A_1 \in \Aset_1, \A_2 \in \Aset_2, \A_1 \Compose \A_2 \neq \bot }.
\]

Lemma~\ref{lem:compose:model-composition} implies that $\phiabstkw_\S$ is a homomorphism from formulas and $*$ to sets of AMS and $\Compose$:

\begin{lemma}\label{lem:compose:correct}
  For all $\phi_1,\phi_2$,
  $\phiabsts{\S}{\phi_1 * \phi_2} = \phiabsts{\S}{\phi_1} \Compose
  \phiabsts{\S}{\phi_2}$.
\end{lemma}
\begin{AppendixOnly}
  \begin{appendixproof}
    \emph{(Lemma~\ref{lem:compose:correct}.)}
    Let $\A \in \phiabsts{\S}{\phi_1 * \phi_2}$.
    There then exists a heap $\H$ such that
    $\SH \ourmodels \phi_1*\phi_2$ and $\mabst\SH=\A$.
    By the semantics of $*$, we can split $\H$ into
    $\H_1\oursunion\H_2$ with $\SHi{i}\ourmodels\phi_i$ (and thus
    $\mabst\SHi{i} \in \phiabsts{\S}{\phi_i}$).
    By Lemma~\ref{lem:compose:model-composition},
    $\A = \mabst\SHi{1}\Compose\mabst\SHi{2}$ for $\H_1$, $\H_2$ as
    above. Consequently,
    $\A \in \phiabsts{\S}{\phi_1} \Compose \phiabsts{\S}{\phi_2}$ by
    definition of $\Compose$.

    Conversely, let
    $\A \in \phiabsts{\S}{\phi_1} \Compose \phiabsts{\S}{\phi_2}$. By
    definition of $\Compose$, there then exist
    $\A_i \in \phiabsts{\S}{\phi_i}$ such that
    $\A=\A_1\Compose\A_2$. Let $\H_1,\H_2$ be witnesses of that, i.e.,
    $\SHi{i}\ourmodels\phi_i$ with $\mabst\SHi{i}=\A_i$. Assume
    w.l.o.g.~that $\H_1\oursunion\H_2\neq\bot$. (Otherwise, replace
    $\H_2$ with an $\H_2'$ such that $\SHi{2} \Iso \SHpair{\S}{\H_2'}$  and   $\H_1\oursunion\H_2'\neq\bot$;
    by Lemma~\ref{lem:iso-same-sat} we then have $\SHpair{\S}{\H_2'}\ourmodels\phi_2$.)
    By the semantics of $*$,
    $\SHpair{\S}{\H_1\oursunion\H_2}\ourmodels \phi_1*\phi_2$.
    Therefore, $\mabst\SHpair{\S}{\H_1\oursunion\H_2} \in \phiabsts{\S}{\phi_1 * \phi_2}$.
    By Lemma~\ref{lem:compose:model-composition},  $\mabst\SHpair{\S}{\H_1\oursunion\H_2} = \A$.
    The claim follows.
  \end{appendixproof}
\end{AppendixOnly}

\paragraph*{The septraction operator.}

We next define an \emph{abstract septraction operator} $\gsept$ that relates to $\Compose$ in the same way that $\sept$ relates to $*$.
For two sets of AMS $\Aset_1,\Aset_2$ we set:
\[
  \Aset_1 \gsept \Aset_2 := \{\A \in \AMS \mid \text{
                                there exists } \A_1
      \in \Aset_1 \text{ s.t. }
      \A \Compose\A_1 \in \Aset_2\}
\]

Then, $\phiabstkw_\S$ is a homomorphism from formulas and $\sept$ to sets of AMS and $\gsept$:

\begin{lemma}\label{lem:gsept:correct}
  For all $\phi_1,\phi_2$, $\phiabsts{\S}{\phi_1 \sept \phi_2} = \phiabsts{\S}{\phi_1} \gsept  \phiabsts{\S}{\phi_2}$.
\end{lemma}
\begin{appendixproof}
\emph{(Lemma~\ref{lem:gsept:correct}.)}
Let $\A \in \phiabsts{\S}{\phi_1 \sept \phi_2}$.
Then there exists a model $\SH$ with $\mabst\SH=\A$ and $\SH\ourmodels\phi_1\sept\phi_2$.
Consequently, there exists a heap $\H_1$ such that $\H\oursunion\H_1\neq\bot$, $\SHi{1}\ourmodels\phi_1$ and $\SHpair{\S}{\H\oursunion\H_1}\ourmodels \phi_2$.
By definition of $\phiabstkw_\S$, we then have $\mabst\SHi{1} \in \phiabsts{\S}{\phi_1}$ and $\mabst\SHpair{\S}{\H\oursunion\H_1} \in \phiabsts{\S}{\phi_2}$.
By Lemma \ref{lem:compose:model-composition}, $\mabst\SHpair{\S}{\H\oursunion\H_1} = \mabst\SH \Compose
\mabst\SHi{1}$.
In other words, we have for $\A = \mabst\SH$ and $\A_1 :=
\mabst\SHi{1}$ that $\A_1 \in \phiabsts{\S}{\phi_1}$ and $\A \Compose \A_1 \in \phiabsts{\S}{\phi_2}$.
By definition of $\gsept$, we hence have $\A \in \phiabsts{\S}{\phi_1} \gsept \phiabsts{\S}{\phi_2}$.

Conversely, let $\A \in \phiabsts{\S}{\phi_1} \gsept \phiabsts{\S}{\phi_2}$.
Then there exists an $\A_1 \in \phiabsts{\S}{\phi_1}$ such that $\A\Compose\A_1 \in \phiabsts{\S}{\phi_2}$.
Let $\H,\H_1$ be heaps with $\mabst\SH = \A$, $\mabst\SHi{1}=\A_1$ and $\SHi{1} \ourmodels \phi_1$.
Assume w.l.o.g.~that $\H\oursunion\H_1\neq\bot$.
(Otherwise, replace $\H_1$ with an $\H_1'$ such that $\SHi{1} \Iso \SHpair{\S}{\H_1'}$ and     $\H\oursunion\H_1'\neq\bot$;
by Lemma~\ref{lem:iso-same-sat} we then have $\SHpair{\S}{\H_1'}\ourmodels\phi_1$.)

By Lemma~\ref{lem:compose:model-composition}, we then have  $\mabst\SHpair{\S}{\H\oursunion\H_1} = \A \Compose \A_1$.  By Cor.~\ref{cor:same-abst--same-sat}, this allows us to conclude that $\SHpair{\S}{\H\oursunion\H_1} \ourmodels \phi_2$.
Consequently, $\SH \ourmodels \phi_1\sept\phi_2$, implying  $\A \in \phiabsts{\S}{\phi_1 \sept \phi_2}$.
\end{appendixproof}


\subsection{Refining the Refinement Theorem: Bounding Garbage}\label{sec:garbage-equivalence}

Even though we have now characterized the set $\phiabsts{\S}{\phi}$
for every formula $\phi$, we do not yet have a way to implement AMS computation:
While $\phiabsts{\S}{\phi}$ is finite if $\phi$ is a spatial atom, the set is infinite in general;
see the cases $\phiabsts{\S}{\neg\phi}$ and
$\phiabsts{\S}{\phi_1\sept\phi_2}$.
However, we note that for a fixed stack $\S$ only the garbage-chunk count $\garb$ of an AMS $\AMquadruple \in \phiabsts{\S}{\phi}$ can be of arbitrary size,
while the size of the nodes $\V$, the edges $\E$ and the negative-allocation constraint $\Rset$ is bounded by $\size{\S}$.
Fortunately, to decide the satisfiability of any fixed formula $\phi$, it is \emph{not} necessary to keep track of arbitrarily large garbage-chunk counts.

We introduce the \emph{chunk size} $\csize{\phi}$ of a formula $\phi$, which provides an upper bound on the number of chunks that may be necessary to satisfy and/or falsify the formula;
$\csize{\phi}$ is defined as follows:
\begin{itemize}
\item $\csize{\emp}=\csize{x \pto y}=\csize{\ls(x,y)} = \csize{x=y} = \csize{x\neq y}:= 1$
\item $\csize{\phi * \psi} := \csize{\phi} + \csize{\psi}$
\item $\csize{\phi \sept \psi} := \csize{\psi}$
\item $\csize{\phi \wedge
    \psi} = \csize{\phi \vee \psi} := \max(\csize{\phi},\csize{\psi})$
\item $\csize{\neg \phi} := \csize{\phi}$.
\end{itemize}
Observe that $\csize{\phi}\leq\size{\phi}$ for all $\phi$.
Intuitively, $\csize{\phi}-1$ is an upper bound on the number of times the operation $\oursunion$ needs to be applied when checking whether $\SH \ourmodels \phi$.
For example, let $\psi := x \pto y * ((b \pto c) \sept (\ls(a,c))$.
Then $\csize{\psi}=2$, and to verify that $\psi$ holds in a model that consists of a pointer from $x$ to $y$ and a list segment from $a$ to $b$, it suffices to split this model $\csize{\phi}-1 = 1$ many times using $\oursunion$ (into the pointer and the list segment).

We generalize the refinement theorem, Theorem~\ref{thm:ams-refinement}, to models whose AMS differ in their garbage-chunk count, provided both garbage-chunk counts exceed the chunk size of the formula:

\begin{theorem}[Refined Refinement Theorem]
\label{thm:csize-upper-bound}
  Let $\phi$ be a formula with $\csize{\phi}=k$.
  Let $m\geq k$, $n \geq k$ and let $\SHi{1},\SHi{2}$ be models with $\mabst\SHi{1}=\tuple{\V,\E,\Rset,m}$,
  $\mabst\SHi{2}=\tuple{\V,\E,\Rset,n}$.
  Then, $\SHi{1} \ourmodels \phi$ iff $\SHi{2}\ourmodels \phi$.
\end{theorem}
\begin{AppendixOnly}
  \begin{appendixproof}
  \emph{(Theorem~\ref{thm:csize-upper-bound}.)}
  We proceed by structural induction on $\phi$.
  We only prove that $\SHi{1} \ourmodels \phi$ implies $\SHi{2}\ourmodels \phi$, as the proof of the other direction is very similar.
  \begin{description}
  \item[Case $\emp$, $x=y$, $x\neq y$, $x \pto y$, $\ls(x,y)$.]
      By Lemmas~\ref{lem:phiabst:emp},~\ref{lem:phiabst:eq},~\ref{lem:phiabst:pto} and~\ref{lem:phiabst:ls} we have that the AMS of all models of $\phi$ have a garbage-chunk count of $0$.
      Thus, $\SHi{1}\nourmodels\phi$ and $\SHi{2}\nourmodels\phi$.
  \item[Case $\phi_1*\phi_2$.]
    Assume $\SHi{1} \ourmodels \phi_1*\phi_2$.
    Let $\H_{1,1},\H_{1,2}$ be such that
    $\H_1=\H_{1,1}\oursunion\H_{1,2}$,
    $\SHpair{\S}{\H_{1,1}}\ourmodels\phi_1$, and
    $\SHpair{\S}{\H_{1,2}}\ourmodels\phi_2$.
    Let $\A_1= \tuple{\V_1,\E_1,\Rset_1,m_1} := \mabst(\H_{1,1})$ and
    $\A_2= \langle\V_2,\E_2,\allowbreak\Rset_2,\allowbreak{}m_2
    \rangle := \mabst(\H_{1,2})$.
    Since $k = \csize{\phi_1}+\csize{\phi_2}$, it follows that, either
    $m_1 \geq \csize{\phi_1}$ or $m_2 \geq \csize{\phi_2}$ (or both).
    Assume w.l.o.g.~that $m_1 \geq \csize{\phi_1}$ and let $\A_1' :=
    \tuple{\V_1,\E_1,\Rset_1,n - \min\{\csize{\phi_2},m_2\}}$.
    Observe that $\mabst\SHi{2} = \A_1' \Compose \A_2$.
    There thus exist by Lemma~\ref{lem:decompose} heaps
    $\H_{2,1}, \H_{2,2}$ such that
    $\SHi{2}=\H_{2,1}\oursunion\H_{2,2}$,
    $\mabst\SHpair{\S}{\H_{2,1}}=\A_1'$ and
    $\mabst\SHpair{\S}{\H_{2,1}}=\A_2$.
    As both $m_1 \geq \csize{\phi_1}$ and
    $n - \min\{\csize{\phi_2},m_2\} \ge k - \min\{\csize{\phi_2},m_2\} \ge \csize{\phi_1}$, we have by the
    induction hypothesis for $\phi_1$ that
    $\SHpair{\S}{\H_{2,1}} \ourmodels \phi_1$.  Additionally, we have $\H_{2,2}\ourmodels\phi_2$ by Theorem~\ref{thm:ams-refinement} (for $m_2 < \csize{\phi_2}$) or by the induction hypothesis (for $m_2 \ge \csize{\phi_2}$).
    Consequently, $\SHi{2}\ourmodels\phi_1*\phi_2$.

  %
  \item[Case $\phi_1\sept\phi_2$.]
    Assume $\SHi{1} \ourmodels \phi_1\sept\phi_2$.
    Let $\H_0$ be such that $\SHpair{\S}{\H_0} \ourmodels\phi_1$ and $\SHpair{\S}{\H_1\oursunion\H_0} \ourmodels \phi_2$.
    We can assume w.l.o.g.~that $\H_2\oursunion \H_0 \neq \bot$---if this is not the case, simply replace $\H_0$ with a heap $\H_0'$ with $\SHpair{\S}{\H_0}\Iso\SHpair{\S}{\H_0'}$,
    $\H_1\oursunion \H_0' \neq \bot$ and $\H_2\oursunion \H_0' \neq \bot$;
    then, $\SHpair{\S}{\H_1 \oursunion \H_0'} \ourmodels \phi_2$ by Lemma~\ref{lem:iso-same-sat}.
    We set $\A_2 = \mabst
    \SHpair{\S}{\H_1\oursunion\H_0}$ and
    $\A_2' = \mabst
    \SHpair{\S}{\H_2\oursunion\H_0}$.
    By Lemma~\ref{lem:compose:model-composition} we have $\mabst
    \SHpair{\S}{\H_1\oursunion\H_0} = \mabst \SHpair{\S}{\H_1} \Compose \mabst \SHpair{\S}{\H_0}$ and
    $\mabst
    \SHpair{\S}{\H_2\oursunion\H_0} = \mabst \SHpair{\S}{\H_2} \Compose \mabst \SHpair{\S}{\H_0}$.
    Hence, $\A_2 = \langle \V_2,\E_2,\Rset_2,m' \rangle$ and
    $\A_2' = \langle \V_2,\E_2,\allowbreak\Rset_2,n' \rangle$ for some $\V_2,\E_2,\Rset_2$ and $m',n' \ge k = \csize{\phi_1\sept\phi_2} = \csize{\phi_2}$.
    It thus follows from the induction hypothesis for $\phi_2$ that
    $\SHpair{\S}{\H_0\oursunion\H_2} \ourmodels \phi_2$.
  \item[Case $\phi_1\wedge\phi_2$, $\phi_1\vee\phi_2$.]
    We then have $\SHi{1}\ourmodels \phi_1$ and/or $\SHi{1}\ourmodels \phi_2$.
    By definition of $\csize{\phi_1\wedge\phi_2}$ resp. $\csize{\phi_1\vee\phi_2}$, it follows that
    $n,m \geq \max(\csize{\phi_1},\csize{\phi_2}) \geq \csize{\phi_i}$.
    We therefore conclude from the induction hypothesis that $\SHi{2} \ourmodels \phi_1$ and/or $\SHi{2} \ourmodels \phi_2$.
    Thus, $\SHi{2}\ourmodels\phi_1 \wedge \phi_2$ resp. $\SHi{2}\ourmodels\phi_1 \vee \phi_2$.
  \item[Case $\neg\phi_1$.]
    Assume $\SHi{1}\ourmodels\neg\phi_1$. Consequently,
    $\SHi{1}\nourmodels\phi_1$.
    Since $m,n \geq \csize{\neg\phi_1}=\csize{\phi_1}$, it
    follows by induction that $\SHi{2}\nourmodels\phi_1$.
    Then, $\SHi{2}\ourmodels\neg\phi_1$. \qedhere
  \end{description}
\end{appendixproof}
\end{AppendixOnly}
\begin{SubmissionOnly}
  \begin{proof}
    If $m=n$, the result follows from
  Lemma~\ref{lem:mabst:sound}.
  If $m\neq n$, assume w.l.o.g.~that $m < n$.
  We proceed by structural induction on $\phi$. We illustrate the
  proof by presenting two cases; you can find a full proof in the
  supplementary material.
  \paragraph*{Case $\phi=x \pto y$.} By Lemma~\ref{lem:phiabst:pto},
  the AMS of all models of $x\pto y$ have a garbage-chunk count of
  $0$. Since $m\geq\csize{x\pto y}=1$ and $n > m$, it follows that
  $\SHi{1}\nourmodels x\pto y$ and $\SHi{2}\nourmodels x\pto y$.
  \paragraph*{Case $\phi=\phi_1*\phi_2$.} Assume $\SHi{1} \ourmodels \phi_1*\phi_2$.
    Let $\H_{1,1},\H_{1,2}$ be such that
    $\H_1=\H_{1,1}\oursunion\H_{1,2}$,
    $\SHpair{\S}{\H_{1,1}}\ourmodels\phi_1$, and
    $\SHpair{\S}{\H_{1,2}}\ourmodels\phi_2$.
    Let $\A_1= \tuple{\V_1,\E_1,\Rset_1,m_1} := \mabst(\H_{1,1})$ and
    $\A_2= \langle\V_2,\E_2,\allowbreak\Rset_2,\allowbreak{}m_2
    \rangle := \mabst(\H_{1,2})$.
    Since $k = \csize{\phi_1}+\csize{\phi_2}$, it follows that, either
    $m_1 \geq \csize{\phi_1}$ or $m_2 \geq \csize{\phi_2}$ (or both).
    Assume w.l.o.g.~that $m_1 \geq \csize{\phi_1}$ and let $\A_1' :=
    \tuple{\V_1,\E_1,\Rset_1,n - \min\{\csize{\phi_2},m_2\}}$.
    Observe that $\mabst\SHi{2} = \A_1' \Compose \A_2$.
    There thus exist by Lemma~\ref{lem:decompose} heaps
    $\H_{2,1}, \H_{2,2}$ such that
    $\SHi{2}=\H_{2,1}\oursunion\H_{2,2}$,
    $\mabst\SHpair{\S}{\H_{2,1}}=\A_1'$ and
    $\mabst\SHpair{\S}{\H_{2,1}}=\A_2$.
    As both $m_1 \geq \csize{\phi_1}$ and
    $n - \min\{\csize{\phi_2},m_2\} \ge k - \min\{\csize{\phi_2},m_2\} \ge \csize{\phi_1}$, we have by the
    induction hypothesis for $\phi_1$ that
    $\SHpair{\S}{\H_{2,1}} \ourmodels \phi_1$.  Additionally, we have $\H_{2,2}\ourmodels\phi_2$ by Theorem~\ref{thm:ams-refinement} (for $m_2 < \csize{\phi_2}$) or by the induction hypothesis (for $m_2 \ge \csize{\phi_2}$).
    Consequently, $\SHi{2}\ourmodels\phi_1*\phi_2$.

  \end{proof}
\end{SubmissionOnly}

This implies that $\phi$ is satisfiable over stack $\S$ iff $\phi$ is satisfiable by a heap that contains at most $\csize{\phi}$ garbage chunks:

\begin{corollary}\label{cor:amsk-sufficient}
  Let $\phi$ be an formula with $\csize{\phi}=k$.
  Then $\phi$ is satisfiable over stack $\S$ iff there exists a heap   $\H$ such that (1) $\mabst\SH=(\V,\E,\Rset,\garb)$ for some
  $\garb \leq k$ and (2) $\SH\ourmodels\phi$.
\end{corollary}
\begin{AppendixOnly}
  \begin{appendixproof}
    \emph{(Corollary~\ref{cor:amsk-sufficient}.)}
    Assume $\phi$ is satisfiable and let $\SH$ be a model with
    $\SH \ourmodels \phi$. Let $\A = \AMquadruple := \mabst\SH$.
    If $\garb \leq k$, there is nothing to show.
    Otherwise, let $\A' := \tuple{\V,\E,\Rset,k}$.
    By Lemma~\ref{lem:ams:poly-models}, we can choose a heap $\H'$ with $\mabst\SHprime=\A'$.
    By Theorem~\ref{thm:csize-upper-bound}, $\SHprime \ourmodels \phi$.
  \end{appendixproof}
\end{AppendixOnly}


\subsection{Deciding SSL by AMS Computation}\label{sec:algorithm}
In light of Cor.~\ref{cor:amsk-sufficient}, we can decide the SSL satisfiability problem by means of a function $\abst{\S}{\phi}$ that computes the (finite) intersection of the (possibly infinite) set $\phiabsts{\S}{\phi}$ and the (finite) set $\AMSks := \set{ \AMquadruple \in \AMS\mid \V=\eqclasses{\S} \text{ and } \garb \leq k }$ for $k = \csize{\phi}$.
We define $\abst{\S}{\phi}$ in Fig.~\ref{fig:algorithm}.
For atomic predicates we only need to consider garbage-chunk-count $0$, whereas the cases $*$, $\sept$, $\wedge$ and $\vee$ require \emph{lifting} the bound on the garbage-chunk count from $m$ to $n\geq m$.
\begin{definition}\label{def:bound-lift}
  Let $m,n\in\N$ with $m \leq n$ and let $\A = \AMquadruple \in \AMS$.
  The \emph{bound-lifting} of $\A$ from $m$ to $n$ is
  \[
\lift{m}{n}{\A} :=
\begin{cases}
  \set{\A} & \text{if } \garb < m\\
  \set{\tuple{\V,\E,\Rset,k} \mid m \leq k \leq n } &
  \text{if } \garb = m. \\
\end{cases}
\]
We generalize bound-lifting to sets of AMS:
$\lift{m}{n}{\Aset} := \bigcup_{\A\in\Aset}\lift{m}{n}{\A}$. \qedef{}
\end{definition}
\begin{figure}[tb!]
  \centering
  \input{tablesandfigures/algorithm}
  \vspace{-0.5cm}
  \caption{Computing the abstract memory states of the models of $\phi$ with stack $\S$.
  \vspace{-0.5cm}}
  \label{fig:algorithm}
\end{figure}

As a consequence of Theorem~\ref{thm:csize-upper-bound}, bound-lifting is sound for all $n \geq \csize{\phi}$, i.e.,
\[
    \lift{\csize{\phi}}{n}{\phiabsts{\S}{\phi} \cap \AMS_{\csize{\phi}}}
  = \phiabsts{\S}{\phi} \cap \AMS_{n}.
\]
By combining this observation with the lemmas  characterizing $\phiabstkw_\S$ (Lemmas~\ref{lem:phiabst:emp},\ref{lem:phiabst:eq},\ref{lem:phiabst:pto},
\ref{lem:phiabst:ls},\ref{lem:phiabst:wedge}, \ref{lem:phiabst:vee},\ref{lem:phiabst:neg}, \ref{lem:compose:correct} and \ref{lem:gsept:correct}), we obtain the correctness of $\abst{\S}{\phi}$:
\begin{theorem}\label{lem:algorithm:correct}
  Let $\S$ be a stack and $\phi$ be a formula.
  Then, $\abst{\S}{\phi}=\phiabsts{\S}{\phi}\cap \AMS_{\csize{\phi},\S}$.
\end{theorem}
\begin{AppendixOnly}
\begin{appendixproof}
  \emph{(Theorem~\ref{lem:algorithm:correct}.)}
  We proceed by induction on the structure of $\phi$:
  \begin{description}
  \item[Case $\emp$, $x=y$, $x\neq y$, $x \pto y$, $\ls(x,y)$.]
  By Lemmas~\ref{lem:phiabst:emp},~\ref{lem:phiabst:eq},~\ref{lem:phiabst:pto} and~\ref{lem:phiabst:ls} and the observation that all models of $\phi$ have garbage-chunk count of $0$.

  \item[Case $\phi_1*\phi_2$.]
    By the induction hypotheses, we have for $1 \leq i \leq 2$ that $\abst{\S}{\phi_i}=\phiabsts{\S}{\phi_i}\cap   \AMS_{\csize{\phi_i},\S}$
    Let $\Aset_i :=\lift{\csize{\phi_1}}{\csize{\phi_1 *
        \phi_2}}{\abst{\S}{\phi_i}}$.
    By Theorem~\ref{thm:csize-upper-bound}, it follows that $\Aset_i=\phiabsts{\S}{\phi_i}\cap    \AMS_{\csize{\phi_1* \phi_2},\S}$.
    By Lemma~\ref{lem:compose:correct}, it thus follows that $\Aset_1 \Compose \Aset_2$ contains all AMS in    $\phiabsts{\S}{\phi_1 * \phi_2}$ that can be obtained by composing AMS with a garbage-chunk count of at most $\csize{\phi_i * \phi_2}$.
    Thus, in particular, (1) $\Aset_1 \Compose \Aset_2 \subseteq \phiabsts{\S}{\phi_1 * \phi_2}$ and (2)
    $\Aset_1 \Compose \Aset_2 \supseteq \phiabsts{\S}{\phi_1 * \phi_2} \cap \AMS_{\csize{\phi_1 * \phi_2},\S}$.
    The claim follows.
  \item[Case $\phi_1 \sept \phi_2$.]
    By the induction hypotheses, we have for $1 \leq i \leq 2$ that
    $\abst{\S}{\phi_i}=\phiabsts{\S}{\phi_i}\cap
    \AMS_{\csize{\phi_i},\S}$.
    Let $\Aset_2 :=\lift{\csize{\phi_2}}{\csize{\phi_1*
        \phi_2}}{\abst{\S}{\phi_2}}$.
    By Theorem~\ref{thm:csize-upper-bound}, it follows that
    $\Aset_2=\phiabsts{\S}{\phi_2}\cap \AMS_{\csize{\phi_1 *        \phi_2},\S}$.
    Thus, in particular, $\Aset_2$ contains every AMS in
    $\phiabsts{\S}{\phi_2}$ that can be obtained by composing an AMS
    in $\AMS_{\csize{\phi_1\sept\phi_2}}=\AMS_{\csize{\phi_2}}$ with
    an AMS from $\phiabsts{\S}{\phi_1}\cap
    \AMS_{\csize{\phi_1},\S}$.
    With Lemma~\ref{lem:gsept:correct} we then get that
    $(\abst{\S}{\phi_1} \gsept \Aset_2) \cap
    \AMS_{\csize{\phi_1\sept\phi_2}}$ is precisely the set of AMS
    $\phiabsts{\S}{\phi_1 \sept \phi_2} \cap
    \AMS_{\csize{\phi_1\sept\phi_2}}$.
  \item[Case $\phi_1 \wedge \phi_2$, $\phi_1 \vee \phi_2$.]
    By the induction hypotheses, we have for $1 \leq i \leq 2$ that    $\abst{\S}{\phi_i}=\phiabsts{\S}{\phi_i}\cap
    \AMS_{\csize{\phi_i},\S}$.
    For $1 \leq i \leq 2$, let
    $\Aset_i :=\lift{\csize{\phi_1}}{\csize{\phi}}{\abst{\S}{\phi_i}}$.
    By Theorem~\ref{thm:csize-upper-bound}, we have
    $\Aset_i=\phiabsts{\S}{\phi_i}\cap
    \AMS_{\csize{\phi},\S}$.
    The claim thus follows from Lemma~\ref{lem:phiabst:wedge} resp. Lemma~\ref{lem:phiabst:vee}.
  \item[Case $\neg \phi_1$.]
    By the induction hypothesis, we have that
    $\abst{\S}{\phi_1}=\phiabsts{\S}{\phi_1}\cap
    \AMS_{\csize{\phi_1},\S}$.
    From Lemma~\ref{lem:phiabst:neg}, it follows that
    $\phiabsts{\S}{\neg\phi_1} \cap \AMS_{\csize{\neg\phi_1},\S} =
    \AMS_{\csize{\neg\phi_1},\S} \setminus \phiabsts{\S}{\phi_1} =
    \AMS_{\csize{\neg\phi_1},\S} \setminus (\phiabsts{\S}{\phi_1}\cap
    \AMS_{\csize{\phi_1},\S})$. The claim follows.
    \qedhere
  \end{description}
\end{appendixproof}
\end{AppendixOnly}

\paragraph*{Computability of $\abst{\S}{\phi}$.}
We note that the operators $\Compose,\gmw,\cap,\cup$ and $\setminus$ are all computable as the sets that occur in the definition of $\abst{\S}{\phi}$ are all finite.
It remains to argue that we can compute the set of AMS for all atomic formulas.
This is trivial for $\emp$, (dis-)equalities, and points-to assertions.
For the list-segment predicate, we note that the set $\abst{\S}{\ls(x,y)} = \AbstLists{x}{y} \cap \AMS_{\csize{0},\S}$ can be easily computed as there are only finitely many abstract lists w.r.t. the set of nodes $\V = \eqclasses\S$.
We obtain the following results:

\begin{corollary}\label{cor:abst:computable}
  Let $\S$ be a (finite) stack.
  Then $\abst{\S}{\phi}$ is computable for all formulas $\phi$.
\end{corollary}

\begin{theorem}\label{thm:ssl:sat-decidable}
  Let $\phi \in \SL$ and let $\vec{x} \subseteq \Var$ be a finite set of variables with $\fvs{\phi} \subseteq \vec{x}$.
  It is decidable whether there exists a model $\SH$ with $\dom(\S) = \vec{x}$ and $\SH \ourmodels \phi$.
\end{theorem}
\begin{AppendixOnly}
\begin{appendixproof}
  \emph{(Theorem~\ref{thm:ssl:sat-decidable}.)}
  We consider stacks $\S$ with $\dom(\S) =  \vec{x}$;
  we observe that $\mathbf{C} := \set{\eqclasses{\S} \mid    \dom(\S) \subseteq \vec{x}}$ is finite; and that all stacks $\S,\S'$ with $\eqclasses{\S}=\eqclasses{\S'}$ have the same abstractions by Lemma~\ref{lem:same-eqclasses-same-mabst}.
  Consequently, we can compute the set
  $\set{\abst{\S}{\phi} \mid \dom(\S) \subseteq  \vec{x}}$ by picking for each element $V\in\mathbf{C}$ one stack $\S$ with $\eqclasses{\S}=\V$, and calculating $\abst{\S}{\phi}$ for this stack.
  By Cor.~\ref{cor:abst:computable},
  $\abst{\S}{\phi}$ is computable for every such stack.
  By Theorem~\ref{lem:algorithm:correct} and
  Cor.~\ref{cor:amsk-sufficient}, $\phi$ is satisfiable over stack $\S$ iff $\abst{\S}{\phi}$ is nonempty.
  Putting all this together, we obtain $\phi$ is satisfiable in stacks of size $n$ if and only if any of finitely many computable sets $\abst{\S}{\phi}$ is nonempty.
\end{appendixproof}
\end{AppendixOnly}

\begin{corollary}
\label{cor:entailment}
  $\phi\entailsx\psi$ is decidable for all finite sets of variables $\vec{x} \subseteq \Var$ and $\phi,\psi \in \SL$ with $\fvs{\phi} \subseteq \vec{x}$ and $\fvs{\psi} \subseteq \vec{x}$.
\end{corollary}
\begin{appendixproof}
  \emph{(Corollary~\ref{cor:entailment}.)}
  $\phi\entailsx\psi$ iff $\phi\wedge\neg\psi$ is unsatisfiable w.r.t. $\vec{x}$, which is decidable by Theorem~\ref{thm:ssl:sat-decidable}. 
\end{appendixproof}


\subsection{Complexity of the SSL Satisfiability Problem}\label{sec:complexity}
It is easy to see that the algorithm $\abst{\S}{\phi}$ runs in exponential time.
We conclude this section with a proof that SSL satisfiability and entailment are actually $\PSpace$-complete.

\paragraph*{$\PSpace$-hardness.}
An easy reduction from quantified Boolean formulas (QBF) shows that the SSL satisfiability problem is $\PSpace$-hard.
The reduction is presented in Fig.~\ref{fig:qbf-to-ssl}.  We encode positive literals $x$ by $(x \pto \nil) * \true$ (the heap contains the pointer $x \pto \nil$) and negative literals by $\neg((x \pto \nil) * \true)$ (the heap does not contain the pointer $x\pto\nil$).
The magic wand is used to simulate universals (i.e., to enforce that we consider both the case $x \pto \nil$ and the case $\emp$, setting $x$ both to true and to false).
Analogously, septraction is used to simulate existentials.
Similar reductions can be found (for standard SL) in \cite{calcagno2001computability}.

\begin{figure}[tb!]
  \centering
  \input{tablesandfigures/qbf-reduction}
  \vspace{-0.7cm}
  \caption{Translation $\qbftr{F}$ from closed QBF formula $F$ (in negation normal
    form) to a formula that is satisfiable iff $F$ is true.
    \vspace{-0.5cm}}
  \label{fig:qbf-to-ssl}
\end{figure}

\begin{lemma}\label{lem:sat:pspace-hard}
  The SSL satisfiability problem is $\PSpace$-hard (even without the $\ls$ predicate).
\end{lemma}

Note that this reduction simultaneously proves the $\PSpace$-hardness of SSL model checking:
If $F$ is a QBF formula over variables $x_1,\ldots,x_k$,
then $\qbftr{F}$ is satisfiable iff
$\SHpair{ \set{x_i \mapsto l_i \mid 1 \leq i \leq n} }{\emptyset} \ourmodels \qbftr{F}$ for some locations $\loc_i$ with $\loc_i \neq \loc_j$ for $i \neq j$.

\paragraph*{$\PSpace$-membership.}
For every stack $\S$ and every bound on the garbage-chunk
count of the AMS we consider, it is possible to encode every AMS by a string of polynomial length.

\begin{lemma}\label{lem:ams:size-linear}
  Let $k\in\N$, let $\S$ be a stack and $n:=k+\size{\S}$.
  There exists an injective function
  $\mathsf{encode}\colon\AMSks \to {\set{0,1}}^{*}$
  such that
    \[\size{\mathsf{encode}(\A)}  \in
  \bigO(n \log(n)) \quad \text{ for all } \A \in \AMSks.\]
\end{lemma}
\begin{AppendixOnly}
  \begin{appendixproof}
    \emph{(Lemma~\ref{lem:ams:size-linear}.)}
    Let $\A = \AMquadruple \in \AMSks$.
    Each of the $\size{\S} \leq n$ variables that occur in $A$ can be
    encoded by a logarithmic number of bits.
    Observe that $\size{\V} \leq \size{\S}$, so $\V$ can be encoded by at most $\bigO(n \log(n) + n)$ symbols (using a    constant-length delimiter between the nodes).
    Each of the at most $\size{\V}$ edges can be encoded by
    $\bigO(\log(n))$ bits, encoding the position of the source and
    target nodes in the encoding of $\V$ by $\bigO(\log(n))$ bits each
    and expending another bit to differentiate between $\eqone$ and
    $\geqtwo$ edges.
    $\Rset$ can be encoded like $\V$. Since $\garb \leq k \leq n$, $\garb$
    can be encoded by at most $\log(n)$ bits.
    In total, we thus have an encoding of length $\bigO(n \log(n))$.
  \end{appendixproof}
\end{AppendixOnly}
An enumeration-based implementation of the algorithm in
Fig.~\ref{fig:algorithm} (that has to keep in memory at most one AMS per subformula at any point in the computation) therefore runs in $\PSpace$:

\begin{lemma}\label{lem:ssl:sat-pspace}
  Let $\phi \in \SL$ and let $\vec{x} \subseteq \Var$ be a finite set of variables with $\fvs{\phi} \subseteq \vec{x}$.
  It is decidable in $\PSpace$ (in $\size{\phi}$ and  $\size{\vec{x}}$) whether there exists a model $\SH$ with $\dom(\S) = \vec{x}$ and $\SH \ourmodels \phi$.
\end{lemma}
\begin{AppendixOnly}
  \begin{appendixproof}
    \emph{(Lemma~\ref{lem:ssl:sat-pspace}.)}
    A simple induction on the structure of $\phi$ shows that it is possible to enumerate the set $\abst{\S}{\phi}$ using at most $\size{\phi}$ registers (each storing an AMS).
    The most interesting case is $\phi_1\sept\phi_2$. Assume we can enumerate the sets $\abst{\S}{\phi_1} = \set{\A_1,\ldots,\A_m}$ and $\abst{\S}{\phi_2}=\set{\B_1,\ldots,\B_n}$ in polynomial space.
    We then use a new register in which we successively enumerate all $\A \in  \AMS_{\csize{\phi_1\sept\phi_2},\S}$.
    This is done as follows:
    we enumerate all pairs of AMS $(\A_i,\B_j)$, $1 \leq i \leq m$, $1 \leq j \leq n$;
    we recognize that $\A \in \AMS_{\csize{\phi_1\sept\phi_2},\S}$ iff $\B_j = \A_i\Compose\A$ for any of these pairs $(\A_i,\B_j)$.
    
  \end{appendixproof}
\end{AppendixOnly}

The $\PSpace$-completeness result,
Theorem~\ref{thm:sat:pspace-complete}, follows by combining Lemmas~\ref{lem:sat:pspace-hard} and~\ref{lem:ssl:sat-pspace}.



\section{Program Verification with Strong-Separation Logic}\label{sec:verification}
 \begin{figure}[tb!]
  \centering
  \begin{tabular}{cc}
  \begin{minipage}[b]{0.45\linewidth}
    \AxiomC{}
    \UnaryInfC{$\triple{x \pto z}{x.\fnext := y}{x \pto y}$}
    \DisplayProof{}\\[12pt]
    \AxiomC{}
    \UnaryInfC{$\triple{x \pto z}{\mathsf{free}(x)}{\emp}$}
    \DisplayProof{}\\[12pt]
  \end{minipage}
  &
  \begin{minipage}[b]{0.45\linewidth}
    \AxiomC{}
    \UnaryInfC{$\triple{\emp}{\mathsf{malloc}(x)}{x \pto m}$}
    \DisplayProof{}\\[12pt]
    \AxiomC{}
    \UnaryInfC{$\triple{\emp}{x := y}{x=y}$}
    \DisplayProof{}\\[12pt]
  \end{minipage}
  \vspace{-0.5cm}
  \end{tabular}
  \begin{tabular}{c}
    \AxiomC{}
    \RightLabel{$x$ different from $y$}
    \UnaryInfC{$\triple{y \pto z}{x := y.\fnext}{y \pto z * x = z}$}
    \DisplayProof{}\\[12pt]
    \AxiomC{}
    \RightLabel{$\phi$ is $x = y$ or $x \neq y$}
    \UnaryInfC{$\triple{\emp}{\mathsf{assume}(\phi)}{\phi}$}
    \DisplayProof{}\\[12pt]
  \end{tabular}
  \vspace{-0.2cm}
  \caption{Local proof rules of program statements for forward symbolic execution.}
  \label{fig:atomic-rules}
\end{figure}

\begin{figure}[tb!]
  \centering
  \begin{tabular}{c}
    \AxiomC{$\tripl{P}{c}{Q}$}
    \LeftLabel{Frame rule}
    \RightLabel{$\vec{x} = \modvars{c}$, $\vec{x}'$ fresh}
    \UnaryInfC{$\triple{A * P}{c}{A[\vec{x}'/\vec{x}] * Q}$}
    \DisplayProof{}\\[12pt]
    %
    \AxiomC{$\tripl{P}{c}{Q}$}
    \LeftLabel{Materialization}
    \RightLabel{$Q \ourmodels \neg((x \pto \nil) \sept
                                \true)$, $z$ fresh}
    \UnaryInfC{$\tripl{P}{c}{x \pto z * ((x \pto z) \sept Q)}$}
    \DisplayProof{}\\[12pt]
  \end{tabular}
  \vspace{-0.3cm}
  \caption{The frame and the materialization rule for forward symbolic execution.
  \vspace{-0.3cm}}
  \label{fig:frame-materialization}
\end{figure}

Our main practical motivation behind SSL is to obtain a decidable logic that can be used for fully automatically discharging verification conditions in a Hoare-style verification proof.
Discharging VCs can be automated by calculi that symbolically execute pre-conditions forward resp. post-conditions backward, and then invoking an entailment checker.
Symbolic execution calculi typically either introduce first-order quantifiers or fresh variables in order to deal with updates to the program variables.
We leave the extension of SSL to support for quantifiers for future work and in this paper develop a forward symbolic execution calculus based on fresh variables.

We target the usual Hoare-style setting where a verification engineer annotates the pre- and post-condition of a function and provides loop invariants.
We exemplify two annotated functions in Fig.~\ref{fig:example-progs};
the left function reverses a list and the right function copies a list.
In addition to the program variables, our annotations may contain logical variables (also known as ghost variables);
for example, the annotations of list reverse only contain program variables, while the annotations of list copy also contain the logical variable $u$ (which is assumed to be equal to $x$ in the pre-condition)\footnote{$m$ is a special program variable introduced for modelling $\mathsf{malloc}$.
}.

\paragraph*{A simple heap-manipulating programming language.}
We consider the six program statements $\mathsf{x.next := y}$,
$\mathsf{x := y.next}$ (where $x$ is different from $y$), $\mathsf{free(x)}$, $\mathsf{malloc(x)}$, $\mathsf{x := y}$ and $\mathsf{assume(\phi)}$, where $\phi$ is $x = y$ or $x \neq y$.
We remark that we do not include a statement $\mathsf{x := x.next}$ for ease of exposition;
however, this is w.l.o.g. because $\mathsf{x := x.next}$ can be simulated by the statements $\mathsf{y := x.next; x := y}$ at the expense of introducing an additional program variable $y$.
We specify the semantics of the considered program statements via a small-step operational semantics.
We state the semantics in Fig.~\ref{fig:semantics-atomic-commands},
where we write $\SHpair{\S}{\H} \xrightarrow{c} \SHpair{\S'}{\H'}$, with the meaning that executing $c$ in state $\SHpair{\S}{\H}$ leads to state $\SHpair{\S'}{\H'}$, and $\SHpair{\S}{\H} \xrightarrow{c} \error$, when executing $c$ leads to an error.
Our only non-standard choice is the modelling of the $\mathsf{malloc}$ statement:
we assume a special program variable $m$, which is never referenced by any program statement and only used in the modelling;
the $\mathsf{malloc}$ statement updates the value of the variable $m$ to the target of the newly allocated memory cell;
we include $m$ in order to have a name for the target of the newly allocated memory cell.
We say program statement $c$ is \emph{safe} for a stack-heap pair $\SHpair{\S}{\H}$ if there is no transition $\SHpair{\S}{\H} \xrightarrow{c} \error$.
Given a sequence of program statements $\sequence=c_1\cdots c_k$, we write $\SHpair{\S}{\H} \xrightarrow{\sequence} \SHpair{\S'}{\H'}$, if there are stack-heap pairs $\SHpair{\S_i}{\H_i}$, with $\SHpair{\S_0}{\H_0} = \SHpair{\S}{\H}$, $\SHpair{\S_k}{\H_k} = \SHpair{\S'}{\H'}$ and $\SHpair{\S_i}{\H_i} \xrightarrow{c_i} \SHpair{\S_{i+1}}{\H_{i+1}}$ for all $1 \le i \le k$.

\begin{figure}[tb!]
  \centering
\begin{tabular}{|l|l|}
  \hline
  $\SHpair{\S}{\H} \xrightarrow{x.\fnext := y} \SHpair{\S'}{\H'}$ & if $\S(x) \in \dom(\H)$, with $\S' = \S$\\
  & \hspace{0.7cm} and $\H' = \H[\S(x)/\S(y)]$ \\
  $\SHpair{\S}{\H} \xrightarrow{x.\fnext := y} \error$ & if $\S(x) \not\in \dom(\H)$ \\      \hline
  $\SHpair{\S}{\H} \xrightarrow{x := y.\fnext} \SHpair{\S'}{\H'}$ & if $\S(y) \in \dom(\H)$, with $\S' = \S[x/\H(\S(y))]$\\
  & \hspace{0.7cm} and $\H' = \H$\\
  $\SHpair{\S}{\H} \xrightarrow{x := y.\fnext} \error$ & if $\S(y) \not\in \dom(\H)$ \\      \hline
  $\SHpair{\S}{\H} \xrightarrow{\mathsf{free}(x)} \SHpair{\S'}{\H'}$ & if $\S(x) \in \dom(\H)$, with $\S' = \S$\\
  & \hspace{0.7cm} and $\H' = \H[\S(x)/\bot]$\\
  $\SHpair{\S}{\H} \xrightarrow{\mathsf{free}(x)} \error$ & if $\S(x) \not\in \dom(\H)$ \\      \hline
  $\SHpair{\S}{\H} \xrightarrow{\mathsf{malloc}(x)} \SHpair{\S'}{\H'}$ & with $\S' = \S[x/l][m/k]$ and $\H' = \H[l/k]$\\
  & \hspace{0.7cm}  for some $l \in \Loc \setminus \dom(\H)$, $k \in \Loc$\\
  \hline
  $\SHpair{\S}{\H} \xrightarrow{x := y} \SHpair{\S'}{\H'}$ & with $\S' = \S[x/\S(y)]$ and $\H' = \H$\\
  \hline
  $\SHpair{\S}{\H} \xrightarrow{\mathsf{assume}(\phi)} \SHpair{\S'}{\H'}$,  &  if $\S(x) = \S(y)$ resp. $\S(x) \neq \S(y)$, \\
  \hspace{0.7cm} where $\phi$ is $x = y$ or $x \neq y$ & \hspace{0.7cm} with $\S' = \S$ and $\H' = \H$\\
  \hline
\end{tabular}
  \vspace{-0.3cm}
  \caption{Semantics of program statements.
  \vspace{-0.3cm}}
  \label{fig:semantics-atomic-commands}
\end{figure}

\paragraph*{Forward Symbolic Execution Rules.}
The rules for the program statements in Fig.~\ref{fig:atomic-rules} are local in the sense that they only deal with a single pointer or the empty heap.
The rules in Fig.~\ref{fig:frame-materialization} are the main rules of our forward symbolic execution calculus.
The frame rule is essential for lifting the local proof rules to larger heaps.
The materialization rule ensures that the  frame rule can be applied whenever the pre-condition of a local proof rule can be met.
We now give more details.
For a sequence of program statements $\sequence=c_1\cdots c_k$ and a pre-condition $P_\mathit{start}$, the symbolic execution calculus derives triples $\tripl{P_\mathit{start}}{c_1\cdots c_i}{Q_i}$ for all $1\le i \le k$.
In order to proceed from $i$ to $i+1$, either  1) only the frame rule is applied or 2) the materialization rule is applied first followed by an application of the frame rule.
The frame rule can be applied if the formula $Q_i$ has the shape $Q_i = A*P$, where $A$ is suitably chosen and $P$ is the pre-condition of the local proof rule for statement $c_i$.
Then, $Q_{i+1}$ is given by $Q_{i+1} = A[\vec{x}'/\vec{x}] * Q$, where $\vec{x} = \modvars{c}$, $\vec{x}'$ are fresh copies of the variables $\vec{x}$ and $Q$ is the right hand side of the local proof rule for statement $c_i$, i.e., we have $\tripl{P}{c_i}{Q}$.
Note that the frame rule requires substituting the modified program variables with fresh copies:
We set $\modvars{c} := \{x,m\}$ for
$c=\mathsf{malloc}(x)$, $\modvars{c} := \{x\}$ for $c = \mathsf{x := y.next}$ and $c = \mathsf{x := y}$, and $\modvars{c} := \emptyset$, otherwise.
The materialization rule may be applied in order to ensure that $Q_i$ has the shape $Q_i = A*P$.
This is not needed in case $P=\emp$ but may be necessary for points-to assertions such as $P = x \pto y$.
We note that $Q_i$ guarantees that a pointer $x$ is allocated iff $Q_i \ourmodels \neg((x \pto \nil) \sept \true)$.
Under this condition, the rule allows introducing a name $z$ for the target of the pointer $x$.
We remark that while backward-symbolic execution calculi typically employ the magic wand, our forward calculus makes use of the dual septraction operator:
this operator allowed us to design a general rule that guarantees a predicate of shape $Q_i = A*P$ without the need of coming up with dedicated rules for, e.g., unfolding list predicates.

\paragraph*{Applying the forward symbolic execution calculus for verification.}
We now explain how the proof rules presented in Fig.~\ref{fig:atomic-rules} and~\ref{fig:frame-materialization} can be used for program verification.
Our goal is to verify that the pre-condition $P$ of a loop-free piece of code $c$ (in our case, a sequence of program statements) implies the post-condition $Q$.
For this, we apply the symbolic execution calculus and derive a triple $\tripl{P}{c}{Q'}$.
It then remains to verify that the final state of the symbolic execution $Q'$ implies the post-condition $Q$.
Here, we face the difficulty that the symbolic execution introduces additional variables:
Let us assume that all annotations are over a set of variables $\vec{x}$, which includes the program variables and the logical variables.
Further assume that the symbolic execution $\tripl{P}{c}{Q'}$ introduced the fresh variables $\vec{y}$.
With the results of Section~\ref{sec:deciding} we can then verify the entailment $Q' \ientails{\vec{x} \cup \vec{y}} Q$.
However, we need to guarantee that all models $\SH$ of $Q$ with $\dom(\S) = \vec{x} \cup \vec{y}$ are also models when we restrict $\dom(\S)$ to $\vec{x}$
(note that the variables $\vec{y}$ are implicitly existentially quantified; we make this statement precise in Lemma~\ref{lem:soundness-symbolic-execution} below).
In order to deal with this issue, we require
annotations to be robust:

\begin{definition}[Robust Formula]
  We call a formula $\phi \in \SL$ \emph{robust}, if for
  all models $\SHpair{\S_1}{\H}$ and $\SHpair{\S_2}{\H}$ with $\fvs{\phi} \subseteq \dom(\S_1)$ and $\fvs{\phi} \subseteq \dom(\S_2)$ and $\S_1(x) = \S_2(x)$ for all $x \in \fvs{\phi}$, we have that $\SHpair{\S_1}{\H} \ourmodels \phi$ iff $\SHpair{\S_2}{\H} \ourmodels \phi$.
\end{definition}

We identify a fragment of robust formulas in the next lemma.
In particular, we obtain that the annotations in Fig.~\ref{fig:example-progs} are robust.

\begin{lemma}
\label{lem:positive-robust}
  Let $\phi \in \SL$ be a positive formula.
  Then, $\phi$ is robust.
\end{lemma}
\begin{appendixproof}
  \emph{(Lemma~\ref{lem:positive-robust}.)}
  Let $\SHpair{\S_1}{\H}$ and $\SHpair{\S_2}{\H}$ be two models with $\S_1(x) = \S_2(x)$ for all $x \in \fvs{\phi}$.
  Then, by Lemma~\ref{lem:positive-same-semantics} we have that $\SHpair{\S_1}{\H} \ourmodels \phi$ iff $\SHpair{\S_1}{\H} \stdmodels \phi$ iff $\SHpair{\S_2}{\H} \stdmodels \phi$ iff $\SHpair{\S_2}{\H} \ourmodels \phi$.
\end{appendixproof}

The following lemma allows us to construct robust formulas from known robust formulas:
\begin{lemma}
\label{lem:robust-builder}
  Let $\phi \in \SL$ be formula.
  If $\phi$ is robust, then $\phi * x \pto y$ and $x \pto y \sept \phi$ are robust.
\end{lemma}
\begin{appendixproof}
  \emph{(Lemma~\ref{lem:robust-builder}.)}
  Immediate from the definition of a robust formula.
\end{appendixproof}

Not all formulas are robust, e.g., consider $\phi$ from Example~\ref{ex:ssl-vs-wsl}.
On the other hand, Lemma~\ref{lem:positive-same-semantics} does not cover all robust formulas, e.g., $\true$ is robust.
We leave the identification of further robust formulas for future work.

\paragraph{Soundness of Forward Symbolic Execution.}
We adapt the notion of a \emph{local action} from~\cite{calcagno2007local} to contracts:
\begin{definition}[Local Contract]
  Given some program statement $c$ and SL formulae $P,Q$, we say the triple $\tripl{P}{c}{Q}$ is a \emph{local contract}, if for every stack-heap pair $\SHpair{\S}{\H}$ with $\SHpair{\S}{\H} \ourmodels P$, every stack $t$ with
  $\S \subseteq t$ and every heap $\H^\circ$ with $\H \oursunionop{t} \H^\circ \neq \bot$, we have that
  \begin{enumerate}[1)]
    \item $c$ is safe for $\SHpair{t}{\H \oursunionop{t} \H^\circ}$, and
    \item for every stack-heap pair $\SHpair{t'}{\H'}$ with $\SHpair{t}{\H \oursunionop{t} \H^\circ} \xrightarrow{c} \SHpair{t'}{\H'}$ there is some heap $\H^{\#}$ with $\H^{\#} \oursunionop{t'} \H^\circ = \H'$ and $\SHpair{t'}{\H^{\#}} \ourmodels Q$.
  \end{enumerate}
\end{definition}
We now state that our local proofs rules specify local contracts:
\begin{lemma}
\label{lem:local-contract}
  Let $c$ be a program statement, and let $\tripl{P}{c}{Q}$ be the triple from its local proof rule as stated in Fig.~\ref{fig:atomic-rules}.
  Then, $\tripl{P}{c}{Q}$ is a local contract.
\end{lemma}
\begin{appendixproof}
  \emph{(Lemma~\ref{lem:local-contract}.)}
  The requirements 1) and 2) of local contracts can be directly verified from the semantics of the program statements.
\end{appendixproof}
We are now ready to state the soundness of our symbolic execution calculus (we assume robust formulas $A$ in the frame rule, which can be ensured by the materialization rule\footnote{Assume that $Q$ is the robust formula currently derived by the symbolic execution, that $c$ is the next program statement, that $\tripl{P_c}{c}{Q_c}$ is the triple from the local proof rule of program statement $c$, and that $Q$ is of shape $Q = A * P_c$.
In case $A$ is not robust (note that this is only possible if $P_c = x \pto z$ for some variables $x$ and $z$), then one can first apply the materialization rule in order to derive formula $Q' = x \pto z * ((x \pto z) \sept Q)$.
Then, $A' = (x \pto z) \sept Q$ is robust by
Lemma~\ref{lem:robust-builder}.
});
we note that the statement makes precise the implicitly existentially quantified variables, stating that there is an extension of the stack to the variables $V$ introduced by the symbolic execution such that $Q$ holds:
\begin{lemma}[Soundness of Forward Symbolic Execution]
\label{lem:soundness-symbolic-execution}
Let $\sequence$ be a sequence of program statements, let $P$ be a robust formula, let $\tripl{P}{\sequence}{Q}$ be the triple obtained from symbolic execution, and let $V$ be the fresh variables introduced during symbolic execution.
Then, $Q$ is robust and for all $\SHpair{\S}{\H}\xrightarrow{\sequence}\SHpair{\S'}{\H'}$ with $\SHpair{\S}{\H} \ourmodels P$, there is a stack $\S''$ with $\S' \subseteq \S''$, $V \subseteq \dom(\S'')$ and $\SHpair{\S''}{\H'} \ourmodels Q$.
\end{lemma}
\begin{appendixproof}
  \emph{(Lemma~\ref{lem:soundness-symbolic-execution}.)}
We prove the claim by induction on the number of applications of the frame rule and materialization rule.
We consider a sequence of program statements $\sequence=c_1\cdots c_n$ for which the triple $\tripl{P}{\sequence}{Q}$ was derived by symbolic execution, introducing some fresh variables $V$.
We then assume that the claim holds for $\tripl{P}{\sequence}{Q}$ (for the base case we allow $\sequence$ to be the empty sequence, i.e, $\sequence = \epsilon$, and consider $\tripl{P}{\epsilon}{P}$) and prove the claim for one more application of the frame rule or the materialization rule.

We first consider an application of the materialization rule.
Then, we have $Q \ourmodels \neg((x \pto \nil) \sept \true)$ (*) and we infer the triple $\tripl{P}{\sequence}{x \pto z * ((x \pto z) \sept Q)}$, where $z$ is some fresh variable.
We now consider some stack-heap pairs $\SHpair{\S}{\H}\xrightarrow{\sequence} \SHpair{\S'}{\H'}$ with $\SHpair{\S}{\H} \ourmodels P$.
Because the claim holds for $\tripl{P}{\sequence}{Q}$ and $V$, there is some stack $\S''$ with $\S' \subseteq \S''$, $V \subseteq \dom(\S'')$ and $\SHpair{\S''}{\H'} \ourmodels Q$.
Because of (*) we have $\SHpair{\S''}{\H'} \ourmodels \neg((x \pto \nil) \sept \true)$.
Hence, there is some $\loc \in \Loc$ such that $\H'(\S''(x)) = \loc$.
We now consider the stack $\S''' = \S''[z/\loc]$.
Note that $\S' \subseteq \S'''$.
Because $Q$ is robust by induction assumption, we then have that $\SHpair{\S'''}{\H'} \ourmodels x \pto z * ((x \pto z) \sept Q)$.
Moreover, we get from Lemma~\ref{lem:robust-builder} that $x \pto z * ((x \pto z) \sept Q)$ is robust.
Thus, the claim is satisfied for the set of variables $V' = V \cup \{z\}$.

We now consider an application of the frame rule, i.e., we need to prove the claim for the sequence $\sequence'=\sequence c$, which extends $\sequence$ by some statement $c$.
Let $\tripl{P_c}{c}{Q_c}$ be the triple from the local proof rule for $c$.
Because the frame rule is applied, we have by assumption that there is some robust SL formula $A$ with $Q = A * P_c$.
From the application of the frame rule we then obtain $\tripl{P}{\sequence'}{A[\vec{x}'/\vec{x}] * Q_c}$, where $\vec{x} = \modvars{c}$ and $\vec{x}'$ fresh.
We now consider some stack-heap pairs $\SHpair{\S}{\H}\xrightarrow{\sequence'} \SHpair{\S'}{\H'}$ with $\SHpair{\S}{\H} \ourmodels P$.
Then, there is some stack-heap pair $\SHpair{\S''}{\H''}$ with
$\SHpair{\S}{\H}\xrightarrow{\sequence} \SHpair{\S''}{\H''}$ and $\SHpair{\S''}{\H''}\xrightarrow{c} \SHpair{\S'}{\H'}$.
Because the claim holds for $\tripl{P}{\sequence}{Q}$ and $V$, there is some stack $t$ with $\S'' \subseteq t$, $V \subseteq \dom(t)$ and $\SHpair{t}{\H''} \ourmodels Q$.
Because of $Q = A * P_c$, we have that there are some heaps $\H_1,\H_2$ with $\H_1 \oursunionop{t} \H_2 = \H''$ such that $\SHpair{t}{\H_1} \ourmodels P_c$
and $\SHpair{t}{\H_2} \ourmodels A$.
Because of $\S'' \subseteq t$ and $\SHpair{\S''}{\H''}\xrightarrow{c} \SHpair{\S'}{\H'}$, we get that
$\SHpair{t}{\H''}\xrightarrow{c} \SHpair{t'}{\H'}$ for some $\S' \subseteq t'$.
Because $\tripl{P_c}{c}{Q_c}$ is a local contract by Lemma~\ref{lem:local-contract}, we get that there is a heap $\H_1'$ with $\H_1' \oursunionop{t'} \H_2 = \H'$ and $\SHpair{t'}{\H_1'} \ourmodels Q_c$.
We now consider the stack $t''$ defined by $t''(x) = t'(x)$ for all $x \in \dom(t')$ and $t''(x') = t(x)$ for all $x \in \modvars{c}$, where $x' \in \vec{x}'$ is the fresh copy created for $x$.
Note that $t' \subseteq t''$.
We recall that $A$ is robust by assumption.
Hence, $\SHpair{t''}{\H_2} \ourmodels A[\vec{x}'/\vec{x}]$.
Moreover, $Q_c$ is robust by Lemma~\ref{lem:positive-robust}.
Hence, $\SHpair{t''}{\H_1} \ourmodels Q_c$.
Thus, $\SHpair{t''}{\H'} \ourmodels A[\vec{x}'/\vec{x}]*Q_c$.
Moreover, we get from Lemma~\ref{lem:robust-builder} that $A[\vec{x}'/\vec{x}]*Q_c$ is robust.
Hence, the claim is satisfied for the set of variables $V' = V \cup \vec{x}'$.
\end{appendixproof}

\begin{figure}[tb!]
\begin{tabular}{r|l}
\begin{minipage}{0.3\linewidth}
\% \emph{list reverse}\\
$\set{\ls(x,\nil)}$\\
\hspace*{0.1cm} $\istmt{a := \nil;}$\\
\hspace*{0.1cm} $\istmt{while(x \neq \nil)}$\\
\hspace*{0.45cm} $\set{\ls(x,\nil) * \ls(a,\nil)}$\\
\hspace*{0.1cm} $\istmt{\{}$\\
\hspace*{0.6cm} $\istmt{b :=  x.next;}$\\
\hspace*{0.6cm} $\istmt{x.next := a;}$\\
\hspace*{0.6cm} $\istmt{a := x;}$\\
\hspace*{0.6cm} $\istmt{w := b;}$\\
\hspace*{0.1cm} $\istmt{\}}$\\
\hspace*{0.1cm} $\istmt{x := w;}$\\
$\set{\ls(x,\nil)}$\\
\end{minipage}
&
\begin{minipage}{0.5\linewidth}
\% \emph{list copy}\\
$\set{\ls(x,\nil) * u = x}$\\
\hspace*{0.1cm} $\istmt{malloc(s);}$\\
\hspace*{0.1cm} $\istmt{r := s;}$\\
\hspace*{0.1cm} $\istmt{while(x \neq \nil)}$\\
\hspace*{0.45cm} $\set{\ls(u,x) * \ls(x,\nil) * \ls(r,s) * s \pto m }$\\
\hspace*{0.1cm} $\istmt{\{}$\\
\hspace*{0.6cm} $\istmt{malloc(t);}$\\
\hspace*{0.88cm} \% \emph{t.data := x.data;  not modelled}\\
\hspace*{0.6cm} $\istmt{s.next := t;}$\\
\hspace*{0.6cm} $\istmt{s := t;}$\\
\hspace*{0.6cm} $\istmt{y := x.next;}$\\
\hspace*{0.6cm} $\istmt{x := y;}$\\
\hspace*{0.1cm} $\istmt{\}}$\\
\hspace*{0.1cm} $\istmt{s.next := \nil;}$\\
$\set{\ls(u,\nil) * \ls(r,\nil)}$\\
\end{minipage}
\end{tabular}
\caption{List reverse (left) and list copy (right) annotated pre- and post-condition and loop invariants.
\vspace{-0.5cm}}
\label{fig:example-progs}
\end{figure}

\paragraph*{Automation.}
We note that the presented approach can fully-automatically verify that the pre-condition of a loop-free piece of code guarantees its post-condition:
For every program statement, we apply its local proof rule using the frame rule (and in addition the materialization rule in case the existence of a pointer target must be guaranteed).
We then discharge the entailment query using our decision procedure from Section~\ref{sec:deciding}.
We now illustrate this approach on the programs from Fig.~\ref{fig:example-progs}.
For both programs we verify that the loop invariant is inductive (in both cases the loop-invariant $P$ is propagated forward through the loop body; it is then checked that the obtained formula $Q$ again implies the loop invariant $P$;
for verifying the implication we apply our decision procedure from Corollary~\ref{cor:entailment}):

\begin{example}
\label{example:reverse}
  Verifying the loop invariant of list reverse:
  \begin{align*}
    &\set{\ls(x,\nil) * \ls(a,\nil)}( =: P)\\
    &\istmt{assume(x\neq\nil)}\\
    &\set{\ls(x,\nil) * \ls(a,\nil) * x \neq \nil}\\
    &\progcmt{materialization}\\
    &\set{x \pto z \sept (\ls(x,\nil) * \ls(a,\nil)* x \neq \nil) * x \pto z}\\
    &\istmt{b := x.next}\\
    &\set{x \pto z \sept (\ls(x,\nil) * \ls(a,\nil)* x \neq \nil) * x \pto z * b = z}\\
    &\istmt{x.next := a}\\
    &\set{x \pto z \sept (\ls(x,\nil) * \ls(a,\nil)* x \neq \nil) * x \pto a * b = z}\\
    &\istmt{a := x}\\
    &\set{x \pto z \sept (\ls(x,\nil) * \ls(a',\nil)* x \neq \nil) * x \pto a' * b = z * a = x}\\
    &\istmt{x := b}\\
    &\, \{x' \pto z \sept (\ls(x',\nil) * \ls(a',\nil) * x' \neq \nil) * x' \pto a' * b = z \ * \\
    & \quad \quad \quad \quad \quad \quad \quad \quad \quad \quad \quad \quad \quad \quad \quad \quad  \quad \quad \quad \quad \quad \quad a = x' * x = b\} ( =: Q)\\
    &\set{\ls(x,\nil) * \ls(a,\nil)}( =: P)
  \end{align*}
\end{example}

\begin{example}
\label{example:copy}
  Verifying the loop invariant of list copy:
  \begin{align*}
    &\set{\ls(u,x) * \ls(x,\nil) * \ls(r,s) * s \pto m}( =: P)\\
    &\istmt{assume(x\neq\nil)}\\
    &\set{\ls(u,x) * \ls(x,\nil) * \ls(r,s) * s \pto m *  x \neq \nil}\\
    &\istmt{malloc(t)}\\
    &\set{\ls(u,x) * \ls(x,\nil) * \ls(r,s) * s \pto m' *  x \neq \nil * t \pto m}\\
    &\istmt{s.next := t}\\
    &\set{\ls(u,x) * \ls(x,\nil) * \ls(r,s) * s \pto t *  x \neq \nil * t \pto m}\\
    &\istmt{s := t}\\
    &\set{\ls(u,x) * \ls(x,\nil) * \ls(r,s') * s' \pto t *  x \neq \nil * t \pto m * s = t}\\
    &\progcmt{materialization}\\
    &\ \{x \pto z \sept (\ls(u,x) * \ls(x,\nil) * \ls(r,s') * s' \pto t *  x \neq \nil * t \pto m * s = t) \ * \\
    & \quad \quad \quad \quad \quad \quad \quad \quad \quad \quad \quad \quad \quad \quad \quad \quad \quad \quad \quad \quad \quad \quad \quad \quad \quad \quad \quad \quad \quad x \pto z\}\\
    &\istmt{y := x.next}\\
    &\, \{x \pto z \sept (\ls(u,x) * \ls(x,\nil) * \ls(r,s') * s' \pto t *  x \neq \nil * t \pto m * s = t) \ * \\
    & \quad \quad \quad \quad \quad \quad \quad \quad \quad \quad \quad \quad \quad \quad \quad \quad \quad \quad \quad \quad \quad \quad \quad \quad \quad \quad x \pto z * y = z\}\\
    &\istmt{x := y}\\
    &\, \{x' \pto z \sept (\ls(u,x') * \ls(x',\nil) * \ls(r,s') * s' \pto t \ *\\
    & \quad \quad \quad \quad \quad \quad \quad \quad x' \neq \nil * t \pto m * s = t) * x' \pto z * y = z * x = y\} ( =: Q)\\
    &\set{\ls(u,x) * \ls(x,\nil) * \ls(r,s) * s \pto m}
    ( =: P)
  \end{align*}
\end{example}

While our decision procedure can automatically discharge the entailments in both of the above examples, we give a short direct argument for the benefit of the reader for the entailment check of Example~\ref{example:reverse} (a direct argument could similarly be worked out for Example~\ref{example:copy}):
We note that $Q'$ simplifies to $Q'' = \set{a \pto x \sept (\ls(a,\nil) * \ls(a',\nil)) * a \pto a' }$.
Every model $\SH$ of $Q''$ must consist of a pointer $a \pto a'$, a list segment $\ls(a',\nil)$  and a heap $\H'$ to which the pointer $a \pto x$ can be added in order to obtain the list segment $\ls(a,\nil)$;
by looking at the semantics of the list segment predicate we see that $\H'$ in fact must be the list segment $\ls(x,\nil)$.
Further, the pointer $a \pto a'$ can be composed with the list segment $\ls(a',\nil)$ in order to obtain $\ls(a,\nil)$.


\section{Normal Forms and the Abduction Problem}\label{sec:abduction}
In this section, we discuss how every AMS can be expressed by a formula, which in turn makes it possible to compute a normal form for every formula.
We then discuss how the normal form transformation has applications to the abduction problem.

\paragraph*{Normal Form.}
We lift the abstraction functions from stacks to sets of variables:
Let $\vec{x} \subseteq \Var$ be a finite set of variables and $\phi \in \SL$ be a formula with $\fvs{\phi} \subseteq \vec{x}$.
We set $\phiabstx{\vec{x}}{\phi} := \set{\phiabsts{\vec{s}}{\phi} \mid \dom(\S) = \vec{x} }$ and $\mabsti{\vec{x}}(\phi) := \phiabstx{\vec{x}}{\phi} \cap \AMS_{\csize{\phi},\vec{x}}$,
where $\AMS_{k,\vec{x}}:= \set{ \AMquadruple \in \AMS\mid \bigcup \V=\vec{x} \text{ and } \garb \leq k }$.
(We note that $\phiabstx{\vec{x}}{\phi}$ is computable by the same argument as in the proof of Theorem~\ref{thm:ssl:sat-decidable}.)

\begin{definition}[Normal Form]
\label{def:formof}
  Let $\normalform{\phi} := \bigvee_{\A \in \phiabstx{\vec{x}}{\phi}} \formof{\csize{\phi}}{\A}$ the \emph{normal form} of $\phi$, where $\formofxm{\A}$ is defined as in Fig.~\ref{fig:formof}.
\qedef
\end{definition}

The definition of $\formofxm{\A}$ represents a
straightforward encoding of the AMS $\A$:
\emph{aliasing} encodes the aliasing between the stack variables as implied by $\V$;
\emph{graph} encodes the points-to assertions and lists of length at least two corresponding to the edges $\E$;
\emph{negalloc} encodes that the negative chunks $R \in \Rset$ precisely allocate the variables $\vec{v} \in R$;
\emph{garbage} ensures that there are either exactly $\garb$ additional non-empty memory chunks that do not allocate any stack variable (if $\garb < m$) or at least $\garb$ such chunks (if $\garb = m$);
\emph{negalloc} and \emph{garbage} use the formula \emph{negchunk} which precisely encodes the definition of a negative chunk.
We have the following result about normal forms:

\begin{theorem}\label{lem:formof:correct}
  $\normalform{\phi} \entailsx \phi$ and $\phi \entailsx \normalform{\phi}$.
\end{theorem}

\begin{figure*}[tb!]
  \begin{align*}
    \formofxm{\A} := & \mathsf{aliasing}(\A)  * \mathsf{graph}(\A) * \mathsf{negalloc}(\A) *
                   \mathsf{garbage}^m(\A)\\
    \mathsf{aliasing}(\A) := & \left(\IteratedStar_{\vec{v} \in   \V, x,y\in \vec{v}}x=y \right) * \left(\IteratedStar_{\vec{v},\vec{w}\in\V, \vec{v} \neq
                               \vec{w}}\max(\vec{v})\neq\max(\vec{w}) \right)\\
    \mathsf{graph}(\A) := & \left(\IteratedStar_{\E(\vec{v})=\tuple{\vec{v}',\eqone}} \max(\vec{v}) \pto \max(\vec{v}')\right) * \\
    & \quad \quad \quad \quad \quad \quad  \left(\IteratedStar_{\E(\vec{v})=\tuple{\vec{v}',\geqtwo}}       \ls_{\geq 2}(\max(\vec{v}), \max(\vec{v}'))\right)\\
    \mathsf{negalloc}(\A) := & \IteratedStar_{R \in \Rset}
        \mathsf{negchunk}(\A) \wedge
        \bigwedge_{\vec{v} \in R}                                \allocp{\max(\vec{v})}
        \wedge
        \bigwedge_{\vec{v} \in V \setminus R} \neg \allocp{\max(\vec{v})}\\
    \mathsf{garbage}^m(\A) :=& \begin{cases}
                               \mathsf{garbage}(\A,\garb) & \text{if }
                               \garb < m \\
                               \mathsf{garbage}(\A,m-1) * \neg \emp \bigwedge_{\vec{v} \in V} \neg \allocp{\max(\vec{v})} & \text{otherwise}\\
                             \end{cases}\\
    \mathsf{garbage}(\A,k) :=&
    \begin{cases}
      \emp & \text{if } k = 0 \\
      \mathsf{garbage}(\A,k-1) * \mathsf{negchunk}(\A) \wedge \bigwedge_{\vec{v} \in V} \neg \allocp{\max(\vec{v})}& \text{otherwise}
    \end{cases}\\
    \mathsf{negchunk}(\A) :=& \neg \emp \wedge \neg(\neg \emp * \neg \emp ) \wedge\\
    & \quad \quad \quad \quad \quad \quad \bigwedge_{\vec{v},\vec{w}\in\V, \phi \in \set{\max(\vec{v}) \pto \max(\vec{w}), \ls(\max(\vec{v}),\max(\vec{w}))}} \neg \phi\\
    \allocp{x} :=& \neg((x \pto \nil) \sept \true) \\
    \ls_{\geq 2}(x,y)  :=& \ls(x,y) \wedge \neg(x \pto y)
  \end{align*}
  \caption{The induced formula $\formofxm{\A}$ of AMS $\A=\AMquadruple$ with $\garb \le m$.}
  \label{fig:formof}
\end{figure*}

\paragraph*{The abduction problem.}
We consider the following relaxation of the entailment problem:
The \emph{abduction problem} is to replace the question mark in the entailment $\phi * [?] \entailsx \psi$ by a formula such that the entailment becomes true.
This problem is central for obtaining a scalable program analyzer as discussed in~\cite{calcagno2011compositional}
\footnote{While the program analyzer proposed in~\cite{calcagno2011compositional} employs \emph{bi-abductive} reasoning, the suggested bi-abduction procedure in fact proceeds in two separate abduction and frame-inference steps, where the main technical challenge is the abduction step, as frame inference can be incorporated into entailment checking. We believe that the situation for SSL is similar, i.e., solving abduction is the key to implementing a bi-abductive prover for SSL; hence, our focus on the abduction problem.}.
The abduction problem does in general not have a unique solution.
Following~\cite{calcagno2011compositional}, we thus consider optimization versions of the abduction problem, looking for \emph{logically weakest} and \emph{spatially minimal} solutions:

\begin{definition}
  Let $\phi,\psi \in \SL$ and $\vec{x} \subseteq \Var$ be a finite set of variables.
  A formula $\zeta$ is the \emph{weakest} solution to the abduction problem $\phi * [?] \entailsx \psi$ if it holds for all abduction solutions $\zeta'$ that $\zeta'\entailsx\zeta$.
  An abduction solution is $\zeta$ \emph{minimal}, if there is no abduction solution $\zeta'$ with $\zeta \entailsx \zeta' * (\neg\emp)$.
\end{definition}

\begin{lemma}
\label{lem:abduction-solution-characterization}
  Let $\phi,\psi$ be formulas and let $\vec{x} \subseteq \Var$ be a finite set of variables.
  Then, 1) the weakest solution to the abduction problem $\phi * [?] \entailsx \psi$ is given by the formula $\phi \mw \psi$, and the 2) weakest minimal solution is given by the formula $\phi \mw \psi \wedge \neg((\phi \mw \psi) * \neg \emp)$.
\end{lemma}
\begin{appendixproof}
  \emph{(Lemma~\ref{lem:abduction-solution-characterization}.)}
  1) follows directly from the definition of the abduction problem and the semantics of $\mw$.

  For 2), we introduce the shorthand $\zeta := \phi \mw \psi \wedge \neg((\phi \mw \psi) * \neg \emp)$.
  We note that $\zeta \entailsx \phi \mw \psi$, and hence $\zeta$ is a solution to the abduction problem by 1).
  Assume further that there is an abduction solution $\zeta'$ with $\zeta \entailsx \zeta' * (\neg\emp)$.
  By 1) we have $\zeta' \entailsx \phi \mw \psi$.
  Hence, $\zeta \entailsx \phi \mw \psi * (\neg\emp)$.
  However, this contradicts $\zeta \entailsx \neg((\phi \mw \psi) * \neg \emp)$.
  Thus, $\zeta$ is minimal.
  Now, consider another minimal solution $\zeta'$ to the abduction problem.
  By 1), we have $\zeta' \entailsx \phi \mw \psi$.
  Because $\zeta'$ is minimal, we have as above that $\zeta' \entailsx \neg((\phi \mw \psi) * \neg \emp)$.
  Hence, $\zeta' \entailsx \zeta$.
  Thus, $\zeta$ is the weakest minimal solution to the abduction problem.
\end{appendixproof}

We now explain how the normal form has applications to the abduction problem.
According to Lemma~\ref{lem:abduction-solution-characterization}, the best solutions to the abduction problem are given by the formulas $\zeta := \phi \mw \psi$ and $\zeta' := \phi \mw \psi \wedge \neg((\phi \mw \psi) * \neg \emp)$.
While it is great that the existence of these solutions is guaranteed, we a-priori do not have a means to \emph{compute} an explicit representation of these solutions nor to further analyze their structure.
However, the normal form operator allows us to obtain the explicit representations $\normalform{\zeta}$ and $\normalform{\zeta'}$.
We believe that using these explicit representations in a program analyzer or studying their properties is an interesting topic for further research.
Here, we establish one concrete result on solutions to the abduction problem based on normal forms:

\emph{We can compute the weakest resp. the weakest minimal solution to the abduction problem from the positive fragment.}
Observe that among the sub-formulas of $\mathsf{aliasing}$ and $\mathsf{graph}$, only the formula $\ls_{\geq 2}$ is negative.
To be able to use $\ls_{\geq 2}(x,y)$ in a positive formula, we first need to add a new spatial atom $\ls_{\geq 2}(x,y)$ to SSL with the following semantics:
$\ls_{\geq 2}(x,y)$ holds in a model iff the model is a list segment of length at least $2$ from $x$ to $y$.
(The whole development in Sections~\ref{sec:separation} and~\ref{sec:deciding} can be extended by this predicate.)
We can then simplify the formula $\mathsf{graph}(\A)$ in $\formofxm{\A}$ by directly translating edges $\E(\vec{v})=\tuple{\vec{v}',\geqtwo}$ to the atom $\ls_{\geq 2}(\max(\vec{v}),\max(\vec{v}'))$.
Then, $\bigvee_{\AMquadruple \in \phiabstx{\vec{x}}{\zeta} \text{ with } \Rset=\emptyset,\garb=0} \formof{\csize{\phi}}{\A}$ for $\zeta = \phi \mw \psi$ or $\zeta = \phi \mw \psi \wedge \neg((\phi \mw \psi) * \neg \emp)$ is the weakest resp. the weakest minimal solution to the abduction problem from the positive fragment.


\section{Conclusion}\label{sec:conclusion}
We have shown that the satisfiability problem for ``strong'' separation logic with lists is in the same complexity class as the satisfiability problem for standard ``weak'' separation logic without any data structures: $\PSpace$-complete.
This is in stark contrast to the undecidability result for standard (weak) SL semantics, as shown in~\cite{demri2018effects}.

We have demonstrated the potential of SSL for program verification:
1) We have provided symbolic execution rules that, in conjunction with our result on the decidability of entailment, can be used for fully-automatically discharging verification conditions.
2) We have discussed how to compute explicit representations to optimal solutions of the abduction problem.
This constitutes the first work that addresses the abduction problem for a separation logic closed under Boolean operators and the magic wand.

We consider the above results just the first steps in examining strong-separation logic, motivated by the desire to circumvent the undecidability result of~\cite{demri2018effects}.
Future work is concerned with the practical evaluation of our decision procedures, with extending the symbolic execution calculus to a full Hoare logic as well as extending the results of this paper to richer separation logics (SL) such as SL with nested data structures or SL with limited support for arithmetic reasoning.


\bibliographystyle{abbrv}
\bibliography{literature}

\appendix

\section*{Appendix}

\paragraph{Isomorphism.}
SL formulas cannot distinguish isomorphic models:

\begin{definition}
  Let $\SH,\SHpair{\S'}{\H'}$ be models.
  $\SH$ and $\SHpair{\S'}{\H'}$ are \emph{isomorphic}, $\SH \Iso \SHpair{\S'}{\H'}$, if there exists a bijection $\isofun\colon (\locs(\H) \cup \img(\S)) \to (\locs(\H') \cup \img(\S'))$ such that (1) for all $x$, $\S'(x) = \isofun(\S(x))$ and (2) $\H' = \{ \isofun(l) \mapsto \isofun(\H(l)) \mid l \in \dom(\H)\}$. \qedef{}
\end{definition}

\begin{lemma}\label{lem:iso-same-sat}
  Let $\SH, \SHpprime$ be models with $\SH \Iso \SHpair{\S'}{\H'}$ and let $\phi \in \SL$.
  Then $\SH \ourmodels \phi$ iff $\SHpair{\S'}{\H'} \ourmodels \phi$.
\end{lemma}
\begin{proof}
  We prove the claim by induction on the structure of the formula $\phi$.
  Clearly, the claim holds for the base cases $\emp$, $x \pto y$, $\ls(x,y)$, $x = y$ and $x\neq y$.
  Further, the claim immediately follows from the induction assumption for the cases $\phi_1 \wedge \phi_2$, $\phi_1 \vee \phi_2$ and $\neg \phi$.
  It remains to consider the cases $\phi_1 * \phi_2$ and $\phi_1 \sept \phi_2$.
  Let $\SH$ and $\SHpair{\S'}{\H'}$ be two stack-heap pairs with $\SH \Iso \SHpair{\S'}{\H'}$.

  We will show that $\SH \ourmodels \phi_1 * \phi_2$ implies $\SHpair{\S'}{\H'} \ourmodels \phi_1 * \phi_2$; the other direction is completely symmetric.
  We assume that $\SH \ourmodels \phi_1 * \phi_2$.
  Then, there are $\H_1,\H_2$ with $\H_1 \oursunion \H_2 = \H$ and
  $\SHpair{\S}{\H_i} \ourmodels \phi_i$ for $i=1,2$.
  We consider the bijection $\isofun$ that witnesses the isomorphism between $\SH$ and $\SHpair{\S'}{\H'}$.
  Let $\H'_1$ resp. $\H'_2$ be the sub-heap of $\H'$ restricted to $\isofun(\dom(\H_1))$ resp. $\isofun(\dom(\H_2))$.
  It is easy to verify that $\H'_1 \oursunion \H'_2 = \H'$ and $\SHpair{\S}{\H_i} \Iso \SHpair{\S'}{\H'_i}$ for $i=1,2$.
  Hence, we can apply the induction assumption and get that $\SHpair{\S'}{\H'_i} \ourmodels \phi_i$ for $i=1,2$.
  Because of $\H'_1 \oursunion \H'_2 = \H'$ we get $\SHpair{\S'}{\H'} \ourmodels \phi_1 * \phi_2$.

  We will show that $\SH \ourmodels \phi_1 \sept \phi_2$ implies $\SHpair{\S'}{\H'} \ourmodels \phi_1 \sept \phi_2$;
the other direction is completely symmetric.
We assume that $\SH \ourmodels \phi_1 \sept \phi_2$.
Hence there is a heap $\H_0$ with $\SHpair{\S}{\H_0} \ourmodels \phi_1$ and $\SHpair{\S}{\H_0 \oursunion \H} \ourmodels \phi_2$.
We note that in particular $\H_0 \oursunion \H \neq \bot$.
We consider the bijection $\isofun$ that witnesses the isomorphism between $\SH$ and $\SHpair{\S'}{\H'}$.
Let $L \subseteq \Loc$ be some subset of  locations with $L \cap (\locs(\H') \cup \img(\S')) = \emptyset$ and $|L| = \locs(\H_0) \setminus (\locs(\H) \cup \img(\S))$.  
We can extend $\isofun$ to some bijective function $\isofun': (\locs(\H_0) \cup \locs(\H) \cup \img(\S)) \to ( L \cup \locs(\H') \cup \img(\S'))$.
Then, $\isofun'$ induces a heap $\H'_0$ such that $\SHpair{\S}{\H_0} \Iso \SHpair{\S'}{\H'_0}$, $\H'_0 \oursunion \H' \neq \bot$ and $\SHpair{\S}{\H_0 \oursunion \H} \Iso \SHpair{\S'}{\H'_0 \oursunion \H'}$. 
By induction assumption we get that $\SHpair{\S'}{\H'_0} \ourmodels \phi_1$
and $\SHpair{\S'}{\H'_0 \oursunion \H'} \ourmodels \phi_2$.
Hence, $\SHpair{\S'}{\H'} \ourmodels \phi_1 \sept \phi_2$.
%

\end{proof}

\end{document}